\documentclass[%
preprint,
superscriptaddress,
 amsmath,
 amssymb,
 aps,
floatfix,
]{revtex4-2}

\usepackage[english]{babel}
\usepackage[utf8]{inputenc}
\usepackage{xr-hyper}
\usepackage{float}

\usepackage{dcolumn}
\usepackage{bm}

\usepackage[colorlinks]{hyperref}
\usepackage[T1]{fontenc}
\usepackage[dvips]{graphicx}
\usepackage[dvipsnames]{xcolor}
\usepackage[font=small,labelfont=bf,justification=justified,format=plain,singlelinecheck=false]{caption}
\usepackage{url}
\usepackage{ulem}
\usepackage{braket}
\usepackage{array}

\usepackage{tikzducks}

\makeatletter
\newcommand*{\addFileDependency}[1]{
  \typeout{(#1)}
  \@addtofilelist{#1}
  \IfFileExists{#1}{}{\typeout{No file #1.}}
}
\makeatother

\newcommand{\abs}[1]{|#1|}

\usepackage{subcaption}

\newcommand{\BISE}{Bi$_2$Se$_3$}

\newcommand{\WSe}[1]{WSe$_{#1}$}

\newcommand{\centeredtxt}[1]{
    \begin{tabular}{l}
        \parbox{2cm}{\vspace{-80pt} \centering #1}
    \end{tabular}
}

\newcommand{\centeredtxta}[1]{
    \begin{tabular}{l}
        \parbox{2cm}{\vspace{-120pt} \centering #1}
    \end{tabular}
}

\newcommand{\centeredtxtaa}[1]{
    \begin{tabular}{l}
        \parbox{2cm}{\vspace{-110pt} \centering #1}
    \end{tabular}
}

\newcommand*{\myexternaldocument}[1]{%
    \externaldocument{#1}%
    \addFileDependency{#1.tex}%
    \addFileDependency{#1.aux}%
}

\myexternaldocument{suppl_mat}

\hypersetup{filecolor=blue}
\hypersetup{citecolor=blue}
\hypersetup{urlcolor=blue}
\hypersetup{linkcolor=blue}

\begin{document}


\title{Multi-functional Wafer-Scale Van der Waals Heterostructures and Polymorphs\\}

\author{M.~Mičica}
\affiliation{Laboratoire de Physique de l’Ecole Normale Supérieure, ENS, Université PSL, CNRS, Sorbonne Université,
Université Paris Cité, F-75005 Paris, France}

\author{A.~Wright}
\affiliation{Laboratoire de Physique de l’Ecole Normale Supérieure, ENS, Université PSL, CNRS, Sorbonne Université,
Université Paris Cité, F-75005 Paris, France}

\author{S.~Massabeau}
\affiliation{Laboratoire Albert Fert, CNRS, Thales, Université Paris-Saclay, 91767 Palaiseau, France}

\author{S.~Ayari}
\affiliation{De Vinci Higher Education, Research Center, 92916 Paris La Défense, France}
\affiliation{Laboratoire de Physique de l’Ecole Normale Supérieure, ENS, Université PSL, CNRS, Sorbonne Université,
Université Paris Cité, F-75005 Paris, France}
\author{E.~Rongione}
\affiliation{Laboratoire de Physique de l’Ecole Normale Supérieure, ENS, Université PSL, CNRS, Sorbonne Université,
Université Paris Cité, F-75005 Paris, France}
\affiliation{Laboratoire Albert Fert, CNRS, Thales, Université Paris-Saclay, 91767 Palaiseau, France}

\author{M.~Oliveira Ribeiro}
\affiliation{Univ.~Grenoble Alpes, CEA, CNRS, Grenoble INP, IRIG-SPINTEC, F-38000 Grenoble, France}

\author{S.~Husain}
\affiliation{Laboratoire Albert Fert, CNRS, Thales, Université Paris-Saclay, 91767 Palaiseau, France}

\author{T.~Denneulin}
\author{R.~Dunin-Borkowski}
\affiliation{Ernst Ruska-Centre for Microscopy and Spectroscopy with Electrons, Forschungszentrum Jülich, D-52425 Jülich, Germany}

\author{J.~Mangeney}
\affiliation{Laboratoire de Physique de l’Ecole Normale Supérieure, ENS, Université PSL, CNRS, Sorbonne Université,
Université Paris Cité, F-75005 Paris, France}

\author{J.~Tignon}
\affiliation{Laboratoire de Physique de l’Ecole Normale Supérieure, ENS, Université PSL, CNRS, Sorbonne Université,
Université Paris Cité, F-75005 Paris, France}

\author{R. Lebrun}
\affiliation{Laboratoire Albert Fert, CNRS, Thales, Université Paris-Saclay, 91767 Palaiseau, France}

\author{H.~Okuno}
\affiliation{Univ. Grenoble Alpes, CEA, IRIG-MEM, 38000 Grenoble, France}

\author{O.~Boulle}
\author{A.~Marty}
\author{F.~Bonell}
\affiliation{Univ.~Grenoble Alpes, CEA, CNRS, Grenoble INP, IRIG-SPINTEC, F-38000 Grenoble, France}

\author{F. Carosella}
\affiliation{Laboratoire de Physique de l’Ecole Normale Supérieure, ENS, Université PSL, CNRS, Sorbonne Université,
Université Paris Cité, F-75005 Paris, France}

\author{H.~Jaffrès}
\affiliation{Laboratoire Albert Fert, CNRS, Thales, Université Paris-Saclay, 91767 Palaiseau, France}

\author{R. Ferreira}
\affiliation{Laboratoire de Physique de l’Ecole Normale Supérieure, ENS, Université PSL, CNRS, Sorbonne Université,
Université Paris Cité, F-75005 Paris, France}

\author{J.-M.~George}
\affiliation{Laboratoire Albert Fert, CNRS, Thales, Université Paris-Saclay, 91767 Palaiseau, France}

\author{M.~Jamet}
\affiliation{Univ.~Grenoble Alpes, CEA, CNRS, Grenoble INP, IRIG-SPINTEC, F-38000 Grenoble, France}

\author{S.~Dhillon}
\affiliation{Laboratoire de Physique de l’Ecole Normale Supérieure, ENS, Université PSL, CNRS, Sorbonne Université,
Université Paris Cité, F-75005 Paris, France}
\affiliation{Corresponding authors:~\href{mailto:matthieu.jamet@cea.fr; sukhdeep.dhillon@phys.ens.fr}{matthieu.jamet@cea.fr; sukhdeep.dhillon@phys.ens.fr}\\
}




\date{\today}

\newpage

\begin{abstract}

Van der Waals heterostructures have promised the realisation of artificial materials with multiple physical phenomena such as giant optical nonlinearities, spin-to-charge interconversion in spintronics and topological carrier protection, in a single layered device through an infinitely diverse set of quantum materials. However, most efforts have only focused on exfoliated material that inherently limits both the dimensions of the materials and the scalability for applications. Here, we show the epitaxial growth of large area heterostructures of topological insulators (\BISE), transition metal dichalcogenides (TMDs, \WSe{2}) and ferromagnets (Co), resulting in the combination of functionalities including tuneable optical nonlinearities, spin-to-charge conversion and magnetic proximity effects. This is demonstrated through coherent phase resolved terahertz currents, bringing novel functionalities beyond those achievable in simple homostructures. In particular, we show the role of different TMD polymorphs, with the simple change of one atomic monolayer of the artificial material stack entirely changing its optical, electrical and magnetic properties. This epitaxial integration of diverse two-dimensional materials offers foundational steps towards diverse perspectives in quantum material engineering, where the material polymorph can be controlled at technological relevant scales for coupling applications in, for example, van der Waals nonlinear optics, optoelectronics, spintronics, multiferroics and coherent current control. 

\end{abstract}


\maketitle

\newpage
\section*{\label{sec:Intro} Introduction}

The potential of van der Waals (vdW) heterostructures~\cite{geim_van_2013,Novoselov2016,Castellanos-Gomez2022,Hou2022,Wang2021} is vast with the possibilities of combining a myriad of physical phenomena such as superconductivity, valleytronics, giant optical nonlinearities, proximity effects and spintronics, in a single artifically created material. It can be achieved through the design and realisation of quantum interfaces and structures by combining materials as diverse as graphene, two-dimensional (2D) transition metal dichalogenides (TMD) and their polymorphs, topological insulators (TIs), 2D ferromagnets, 2D ferroelectrics and many more. This has led to the term of atomic-scale Lego that is facilitated by the absence of dangling bonds in vdW materials. Nevertheless, the vast majority of works that have been performed are based on exfoliation from bulk crystals and manual stacking of individual layers. This inherently inhibits scalability and applications of complex vdW heterostructures. However, recent developments in epitaxial growth of TMDs have shown that large area 2D crystalline materials are starting to become available~\cite{Zhuang22,Li2024,Xue2024}, opening up new possibilities of complex heterostructures. In this work, we show wafer scale epitaxially grown heterostructures of TIs~\cite{Moore2010}, 2D TMD~\cite{Manzeli2017} polymorphs and ultrathin ferromagnets. This opens up exciting prospects of combining multiple effects in the same 2D stack. Here we demonstrate the control of the material's ultrafast transport and terahertz (THz) properties through polymorph dependent optical nonlinearities, ultrafast spin-to-charge conversion (SCC), magnetic proximity effects and anisotropic charge properties over macroscopic surface areas with sensitivity to the monolayer limit.

Regarding the state-of-the-art, over the last decade, layered 2D materials such as TIs~\cite{Moore2010}, Rashba materials~\cite{bihlmayer_rashba-like_2022} and TMDs~\cite{Manzeli2017} and their corresponding quantum interfaces have taken a central role in the range of condensed matter phenomena including, but not limited to: \\
\indent \textit{i}) Spin-to-charge interconversion where the harvesting of strong spin-orbit coupling (SOC) in quantum interfaces has been applied to a diverse range of applications. In particular, SCC in TIs interfaces have demonstrated a rich physics playground owing to their exotic bandstructure properties~\cite{Hasan2010,ZhangZhang2012}, where their electronic band dispersion is composed of a 'bulk' insulating form and Dirac-like topologically-protected conducting surface states within the gap. Recent studies have also investigated ultrafast SCC through THz emission studies where TI/ferromagnet junctions are optically excited to generate a spin current in the ferromagnet and converted into a charge current in the TI~\cite{seifert_efficient_2016, park_ultrafast_2022, wang_ultrafast_2018, tong_enhanced_2020, Rongione23_BiSb,rongione_ultrafast_2022}. Similarly, TMDs with large spin-orbit coupling have been studied, with the layer-by-layer control permitting to engineer and control the SCC processes~\cite{Abdukayumov24}. \\
\indent \textit{ii}) Magnetic proximity effects that have recently emerged, which provide a unique method of engineering 2D materials~\cite{huang_emergent_2020}. Here the inherent compatibility of 2D materials permits a strong interfacial interaction between two entirely different 2D materials i.e non-magnetic and magnetic, modifying each of the layer properties. This permits, for example, to induce magnetic phenomena in the non-magnetic material and has shown, for example, proximity-induced magnetism at the interface between a topological insulator and a ferromagnet, probed using magnetic nonlinearities~\cite{lee_direct_2016}. This has allowed  prospects in the control of magnetic ordering, spintronics and valleytronics without the need of magnetic doping or the use of defects that can negatively impact the crystal structure of 2D materials. \\
\indent \textit{iii}) Giant optical nonlinearities, where a wide range of work has been applied to second harmonic generation (SHG) in 2D TMDs, such as WSe$_2$~\cite{ribeiro-soares_second_2015,Wang15_giantchi2_WSe2}, MoS$_2$~\cite{Day2026}, MoSe$_2$~\cite{Chen2017}, and more~\cite{Lucking2018,Shi2023}. These materials have shown considerable advances with their electronic, optical and nonlinear properties controllable by the number of layers, as well as the layer polymorph of the material. Indeed, the layer polymorph, corresponding to the stacking order of the 2Ds layer has shown to be an important parameter in the control of SHG. This has been demonstrated in, for example, the 2H or 3R stacking in typical TMDs or in ReS$_2$ with distinct differences between 1T$^\prime$ and non-centrosymmetric 2H-polymorphs~\cite{Kucukoz2022}. Note that in SHG measurements, only the intensity is typically measured such that the phase information, and hence the sign of the nonlinear photocurrents and/or polarisation, is lost. \\

\begin{figure}[!h]
  \begin{center}
      \includegraphics[width=1\linewidth]{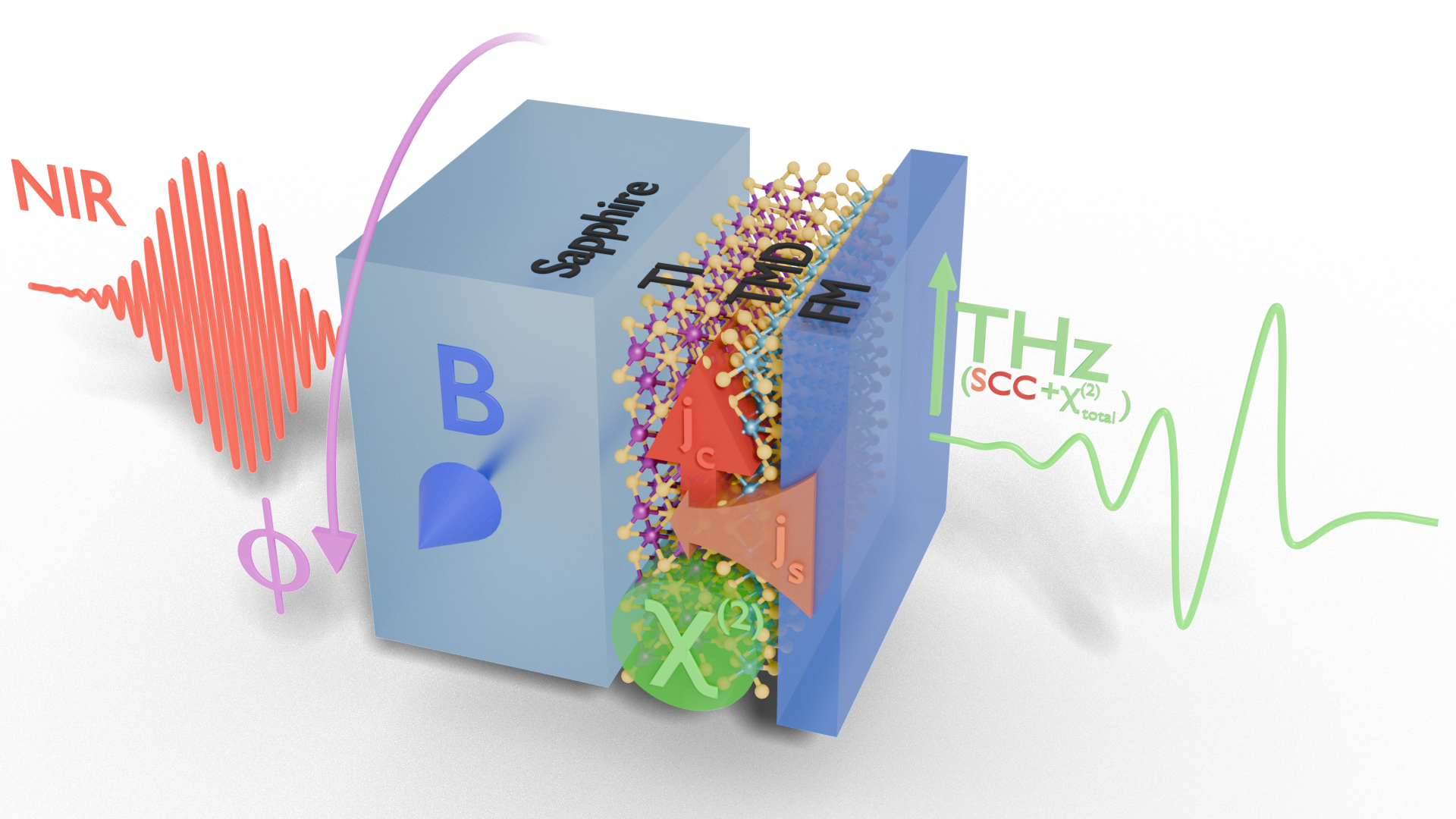}\\\includegraphics[width=0.49\linewidth]{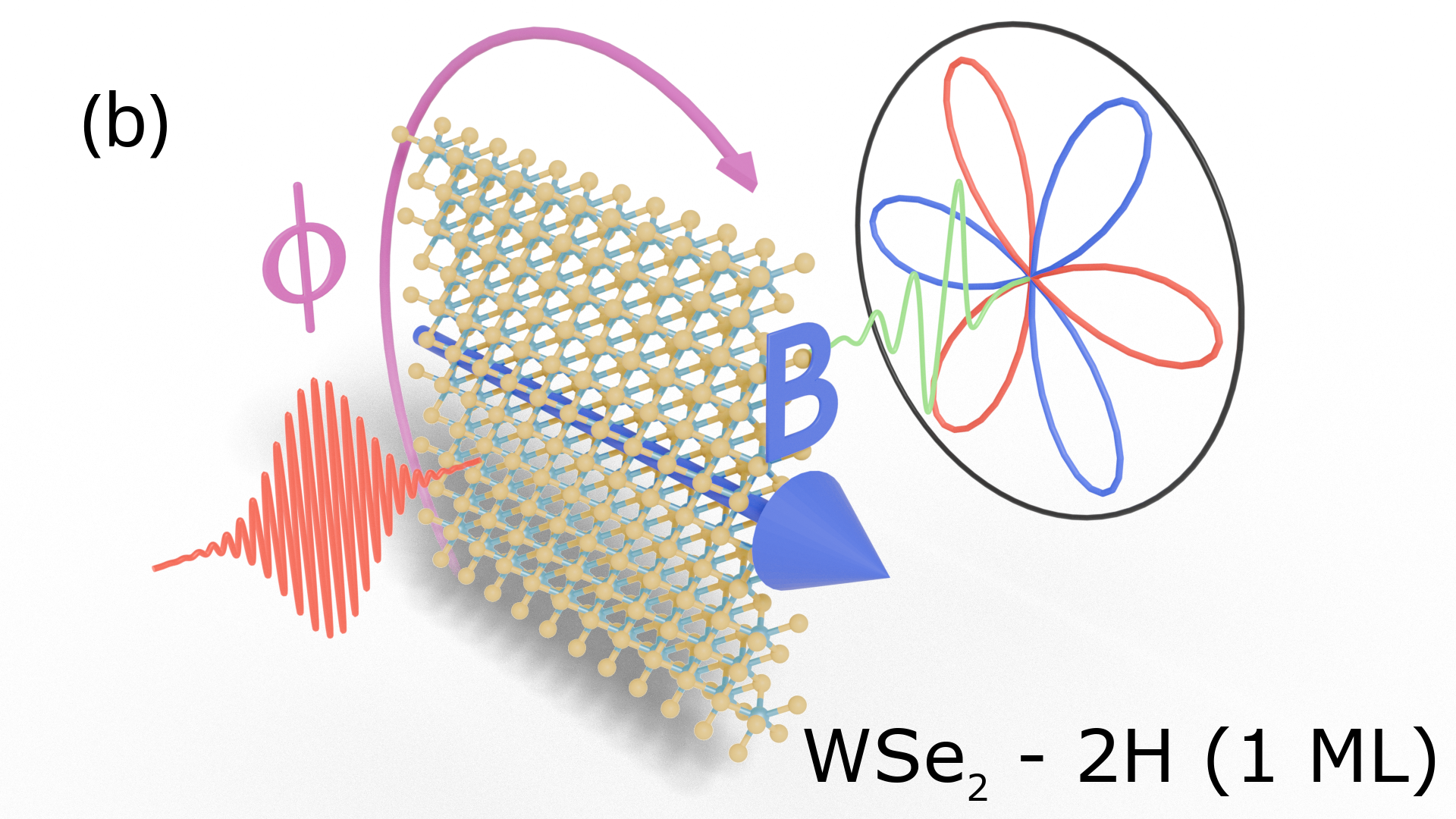}\includegraphics[width=0.49\linewidth]{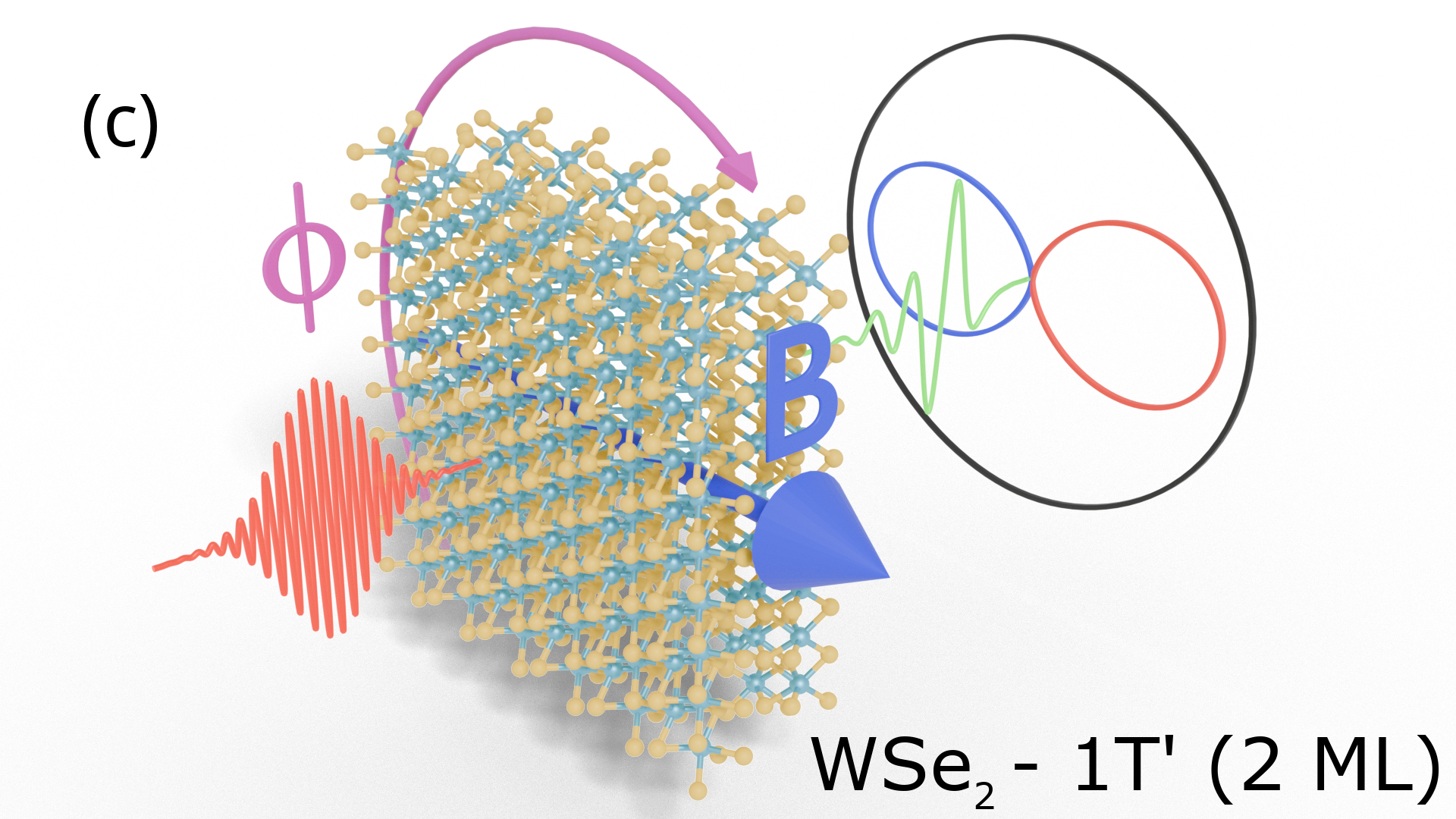}
       \caption{\textbf{Schematic of TI/TMD/FM heterostructure} (a) A femtosecond laser pulse excites the whole structure and generates THz radiation by i) SCC at the TI interface and in the TMD, owing to the flow of spin current $j_s$ from FM and the generation of $j_c$ and ii) by the second order nonlinearities $\chi^{(2)}_{total}$ (see Eq.~\ref{eg:x_decomp}) possessed by both the surface states of the TI and the TMD. The TMD also possesses a magnetic dependent nonlinearity $\chi^{(2)}_{c}$ induced by its proximity with the FM (see Eq.~\ref{eg:x_decomp}). The heterostructure is on a sapphire substrate that is rotated around its azimuthal angle $\phi$ independently of the fixed external magnetic field B. (b-c) Shows an illustration of azimuthal dependence of emitted THz pulse with the different physical contributions, nonlinear (red/blue) and magnetic (black) origin, for different polytypes of \WSe2, 2H and 1T$^\prime$ with 1 ML and 2 ML, respectively.}
    \label{fig_schema}
  \end{center}
\end{figure}

However, the majority of these studies, especially those with TMDs, have been limited to small individual crystals or exfoliated materials, constrained to one single phenomena, and has restricted the study of these materials to the visible/infrared regions of the electromagnetic spectrum. Such effects on large surface areas have been limited owing to the requirement of high quality epitaxial growth of monolayers (MLs). Recent work, however, has shown that the field is progressing rapidly, with demonstrations of polymorph control~\cite{Qin24_3R_epitaxy}, ultrafast photocurrents~\cite{Hemmat2023, huang_surface_2017,pettine_ultrafast_2023} and carrier dynamics~\cite{Yang21,Docherty14}, mid-infrared detection~\cite{yu_atomically_2018}, layered controlled THz spintronic emission~\cite{Abdukayumov24} and mid-infrared harmonic generation~\cite{zhu_ultrastrong_2023,yoshikawa_interband_2019}.

In this work, we show the combination of 2D giant optical nonlinearities, THz spintronics and magnetic proximity-induced effects in epitaxal large area TI/TMD/FM heterostructures. After demonstrating the epitaxial growth of TI/TMD/FM (\BISE/\WSe{2}/Co) with a varying thickness of the WSe$_2$ layers at the atomic level, we report on combined physical phenomena in the 2D stack: \textit{i}) ultrafast spin-to-charge conversion from optically excited spin current in the FM to the TI and the TMD; \textit{ii}) THz nonlinear current generation through optical nonlinearities from the TI and TMD; and \textit{iii}) how the TMD polymorph can be used to control the nonlinearities, the anisotropic response and the proximity-induced magnetic properties at the ML level.

Figure~\ref{fig_schema}(a) shows the schematic of our approach. Here, a femtosecond near infrared laser, $E_{NIR}$, is used to excite a 2D heterostructure of the FM, TMD and TI quantum layers at normal incidence. The FM permits the generation of an ultrafast spin polarised current ($j_s$) that is converted into a charge current ($j_c$) at the surface of TI, whilst both the TMD and TI possess photoexcited nonlinear currents owing to their second order nonlinearities $\chi^{(2)}$. Both these effects result in a radiating electromagnetic pulse through the down-conversion of the optical beam to the THz region $E_{THz}$. Figure~\ref{fig_schema}(b) and (c) shows schematically the effect of isotropic (1 ML) and anisotropic (2 ML) \WSe{2} response, respectively, with the expected azimuthal dependence of the generated THz field from the nonlinearity, where the symmetry of the \WSe{2} plays a crucial role (red/blue curves, see below), and the spin-to-charge conversion (black curve). The \WSe{2} layer, as well as preserving the surface states of the TI ~\cite{galceran_passivation_2022,tong_enhanced_2020,Sun2019, park_ultrafast_2022}, also acts as a tunnel barrier in analogy of spin injection studies on metallic spintronic layers~\cite{hawecker_spin_2021}. 

A key point of this work is on the crystal structure of the TMD structure over large surface areas and its role in the nonlinear and magnetic properties of the heterostructure. Indeed, TMDs are known to have different polymorphs that can effect their optical and electronic response~\cite{Kim24_phase_eng}. Table~\ref{tab_layer} shows the typical polymorphs observed in the TMD \WSe{2} and the expected nonlinear and magnetic response with azimuthal angle. The structural representation of 1T$^\prime$, 2H and 3R TMD polytypes together with their atom coordination are shown. (The numbers indicate the number of layers in the unit cell and the letters stand for trigonal, hexagonal and rhombohedral, respectively. 1T$^\prime$ is a result of the structure distortion of the 1T polymorph). For the 3R polymorph, the structure is non-centrosymmetric and a second order intrinsic nonlinearity $\chi^{(2)}_{i}$ is expected no matter the number of layers~\cite{zeng_controlled_2019}. For the 2H (1T$^\prime$) polymorph, the $\chi^{(2)}_{i}$ is only present for odd (even) number of layers owing to even (odd) layers being centrosymmetric~\cite{zeng_controlled_2019}. The form of the second order polarisation as a function of azimuthal angle, $\phi$ is shown in the 5th row. In the case of 3R and 2H, in the presence of a $\chi^{(2)}_{i}$, the expected form corresponds to a 6-fold rotation symmetry, owing to the 3-fold symmetry of a typical TMD (reduced point group $D_{3h}$ for 2H and $C_{3v}$):
\begin{equation}
    \chi^{(2)}_{i} = d_{26} \cos 3\phi
\end{equation}
where $d$ corresponds to the non-zero nonlinear tensor for this polymorph. However, for the case of 1T$^\prime$, in the presence of a $\chi^{(2)}_{i}$ (even layers), the expected symmetry changes drastically to 2-fold owing to the reduced symmetry of the space group $C_s^1$~\cite{Beams2016}.
\begin{equation}
    \chi^{(2)}_{i} = d_{11} \cos^{3} \phi + (d_{12} + 2d_{26}) \cos \phi  \sin^{2} \phi
\end{equation}
The details of the calculated nonlinear susceptibility using DFT simulations is given in the Supp. Info. The magnetic response (THz emission of spin origin, 6th row) with azimuthal angle is expected to be isotropic at the TMD or TI surface for each of the polymorphs, as the origin is from spin-to-charge conversion (through the Inverse Spin-Hall or Edelstein Effects). However, owing to proximity effects between the ferromagnetic and the TMD layer a small magnetic field nonlinearity is induced with a similar azimuthal response to the nonlinear $\chi^{(2)}_{i}$ response, as shown in Table~\ref{tab_layer}, but with a non-trivial response with number of atomic layers. Indeed the total nonlinearity $\chi^{(2)}_{total}$ is given by:
\begin{equation}\label{eg:x_decomp}
    \chi^{(2)}_{total} = \chi^{(2)}_{i} {\pm} \chi^{(2)}_{c}
\end{equation}
where $\chi^{(2)}_{c}$ is the magnetic nonlinearity, referred to as c-type that reverses sign when the magnetic field is reversed. Typically i-type is orders of magnitude larger than c-type, and therefore the magnetic nonlinearity is often neglected. However, recent work has shown that the c-type nonlinearity can have a important contribution to the total nonlinearity in 2D materials, ~\cite{hou_extraordinary_2024,Toyoda21_SHG_ME_material}.

Note that in such semiconducting TMDs the electronic band structure changes with the layer thickness, with a direct bandgap for ML TMDs to an indirect bandgap for greater number of MLs~\cite{yeh_layer-dependent_2015}. For WSe$_2$ 2H and 3R stacked bilayers, the bandgap energy  is $\simeq$1.8~eV and 1.3~eV $\simeq$ for bulk~\cite{Ramasubramaniam12,He14_stacking}, respectively. For the 1T$^\prime$ polymorph, reports have shown small bandgap energies of around 100~meV~\cite{chen_large_2018,ugeda_observation_2018} for 1 ML or semi-metal behaviour. The bandstructures of each polymorph are presented in the Supp. Info.

\begin{table}[ht]
\begin{tabular}{c||c|c|c}
     Polytype & 2H & 3R & 1T$^\prime$ \\ \hline \hline
     \centeredtxtaa{Top view} &  \includegraphics[width=0.24\linewidth]{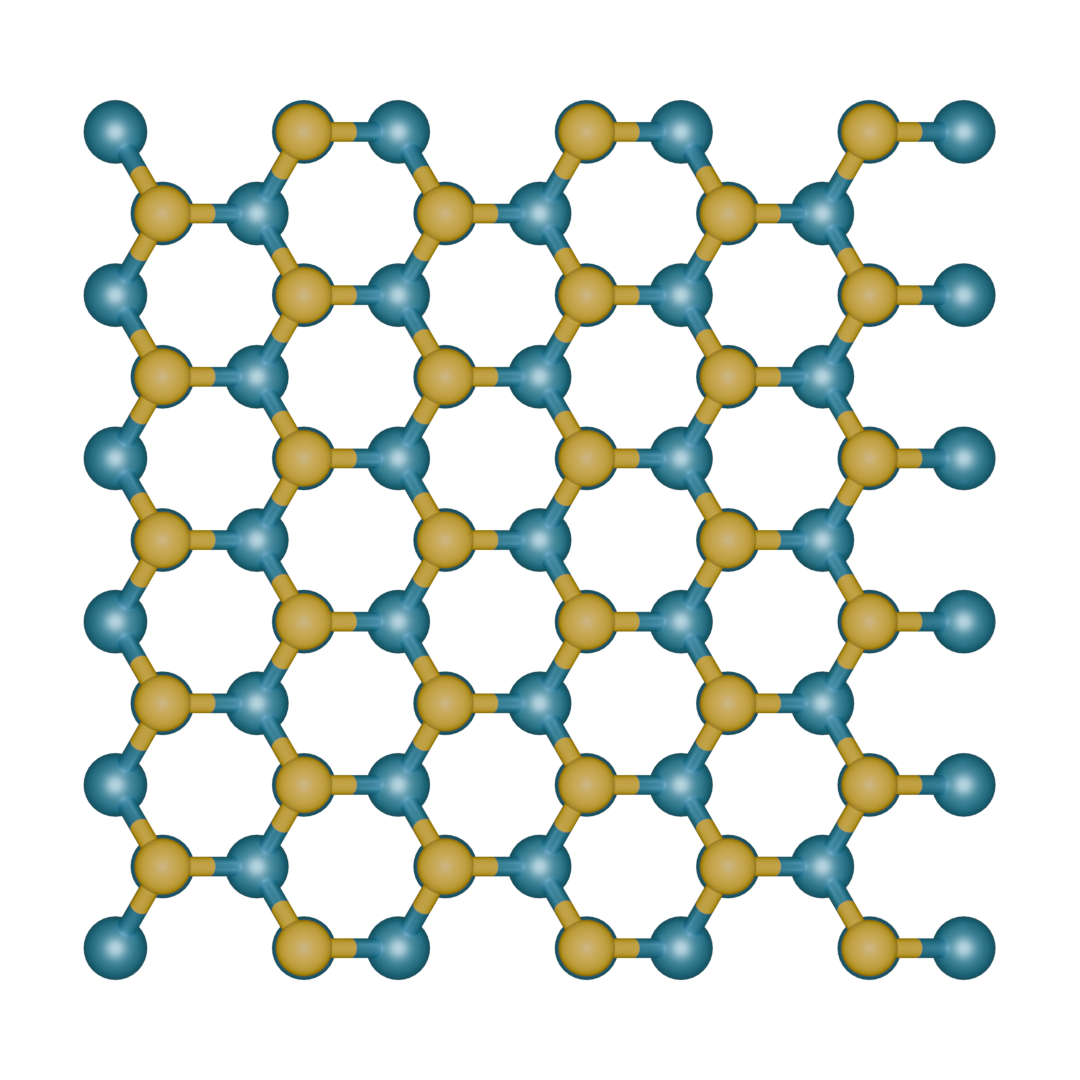} &  \includegraphics[width=0.24\linewidth]{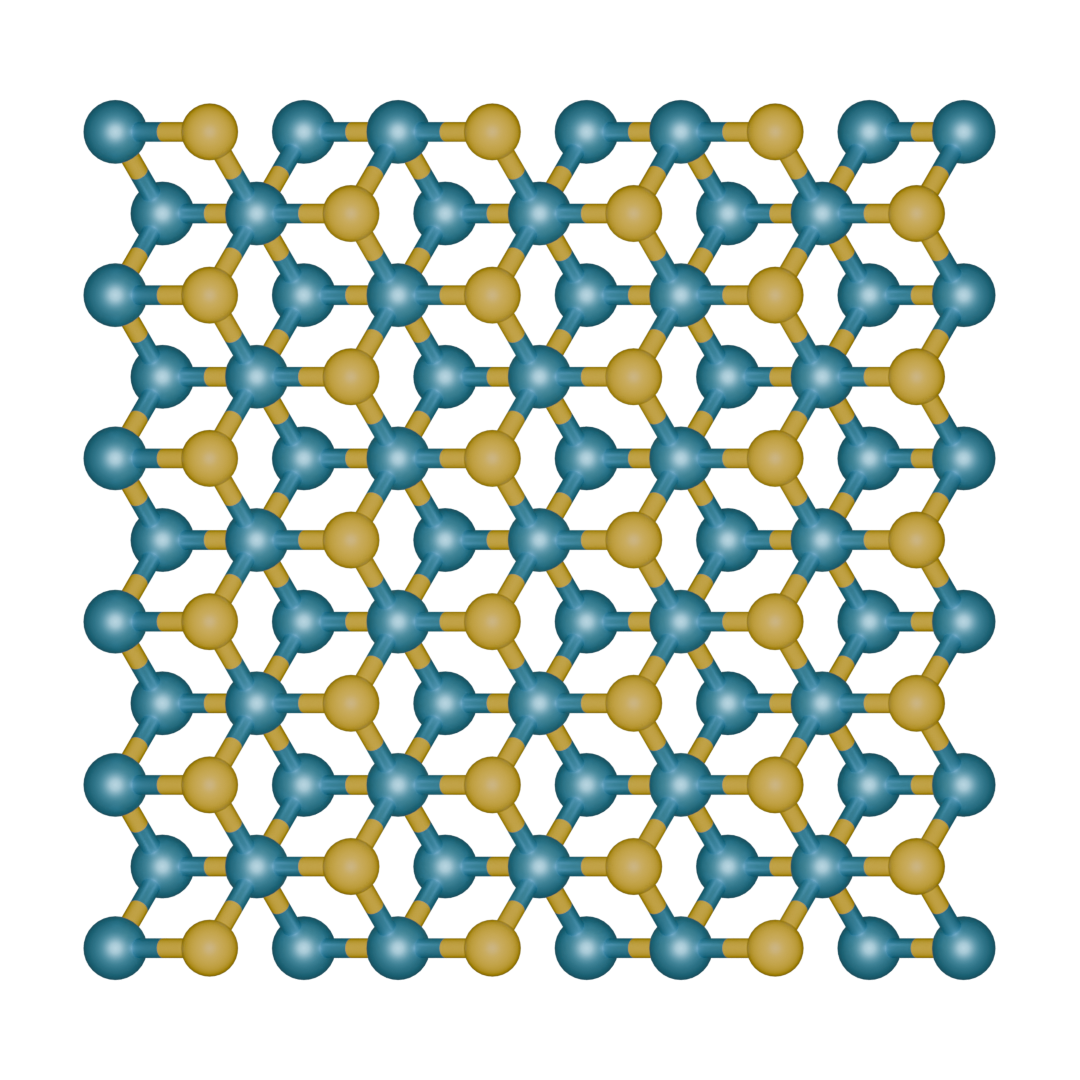} &  \includegraphics[width=0.22\linewidth]{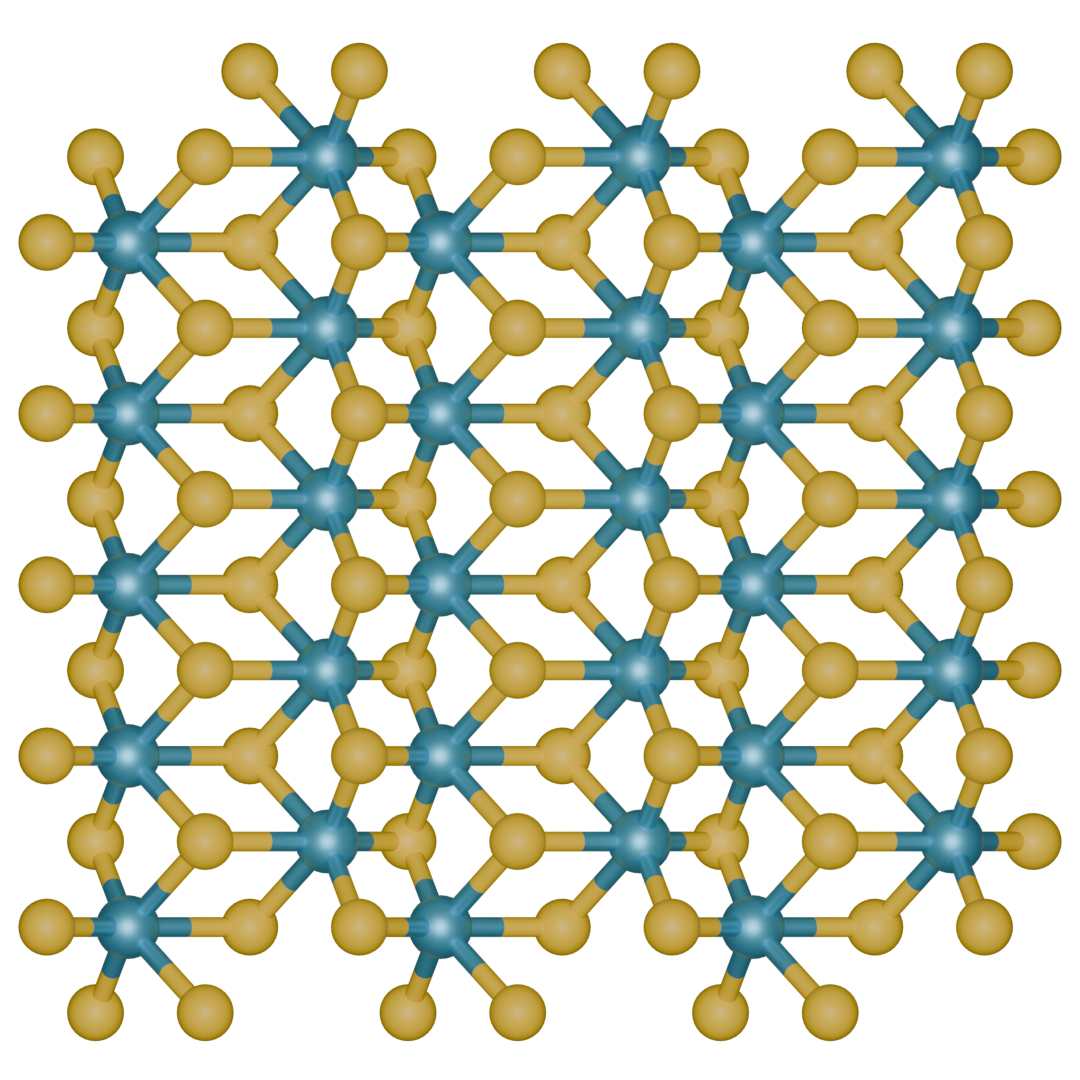} \\ \hline
     \centeredtxt{Side view}& \includegraphics[width=0.26\linewidth]{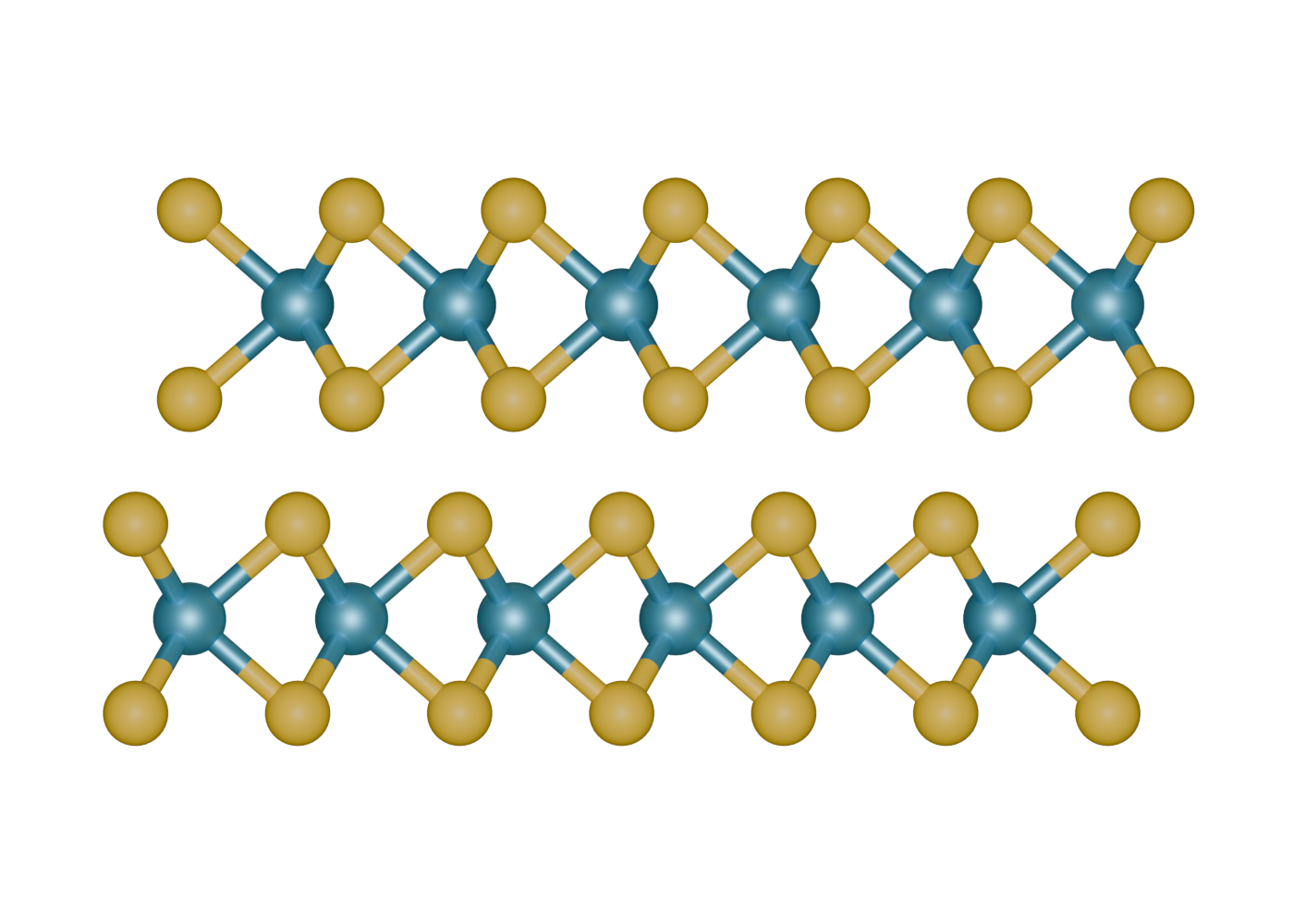}& \includegraphics[width=0.26\linewidth]{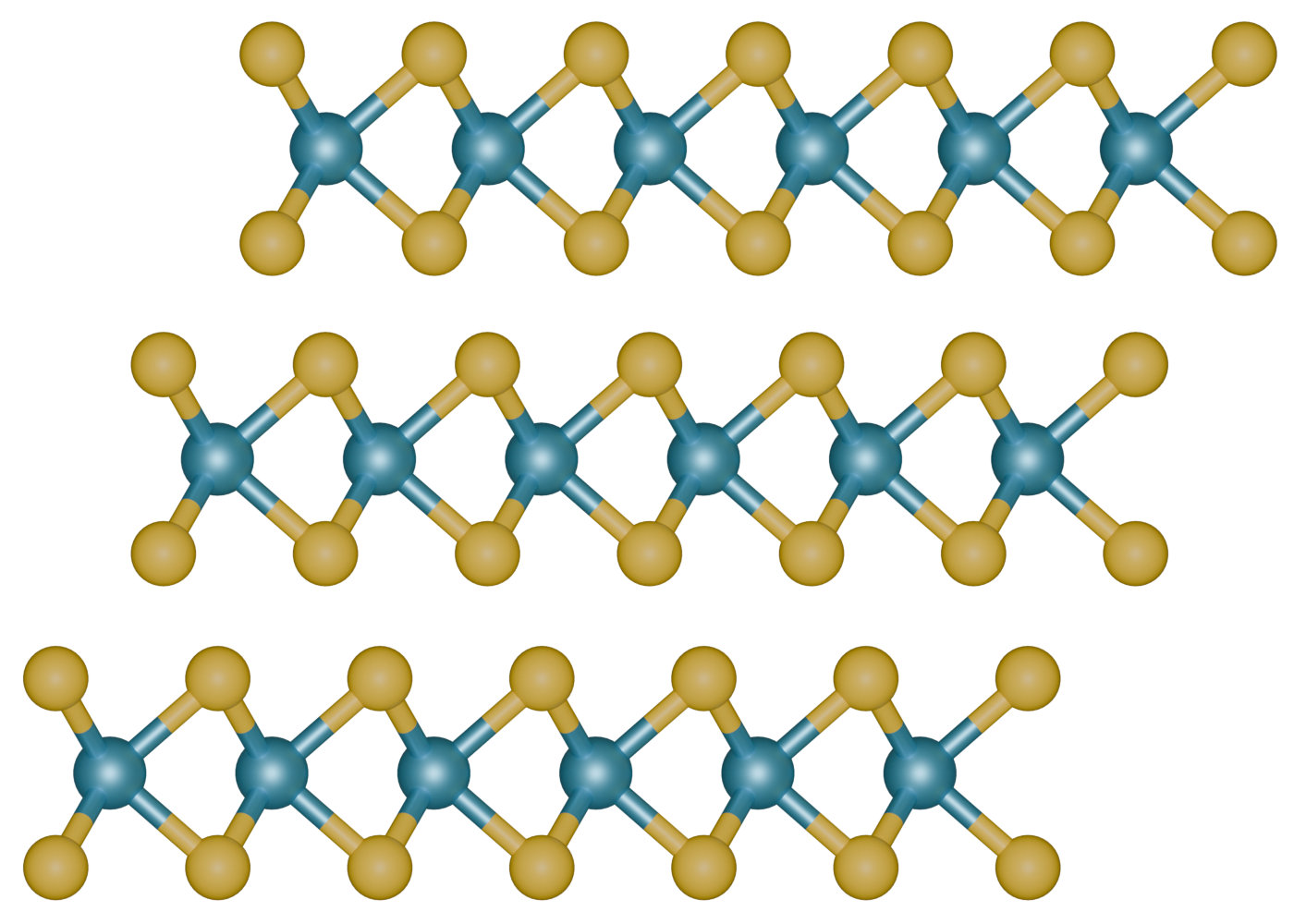}&
     \includegraphics[width=0.26\linewidth]{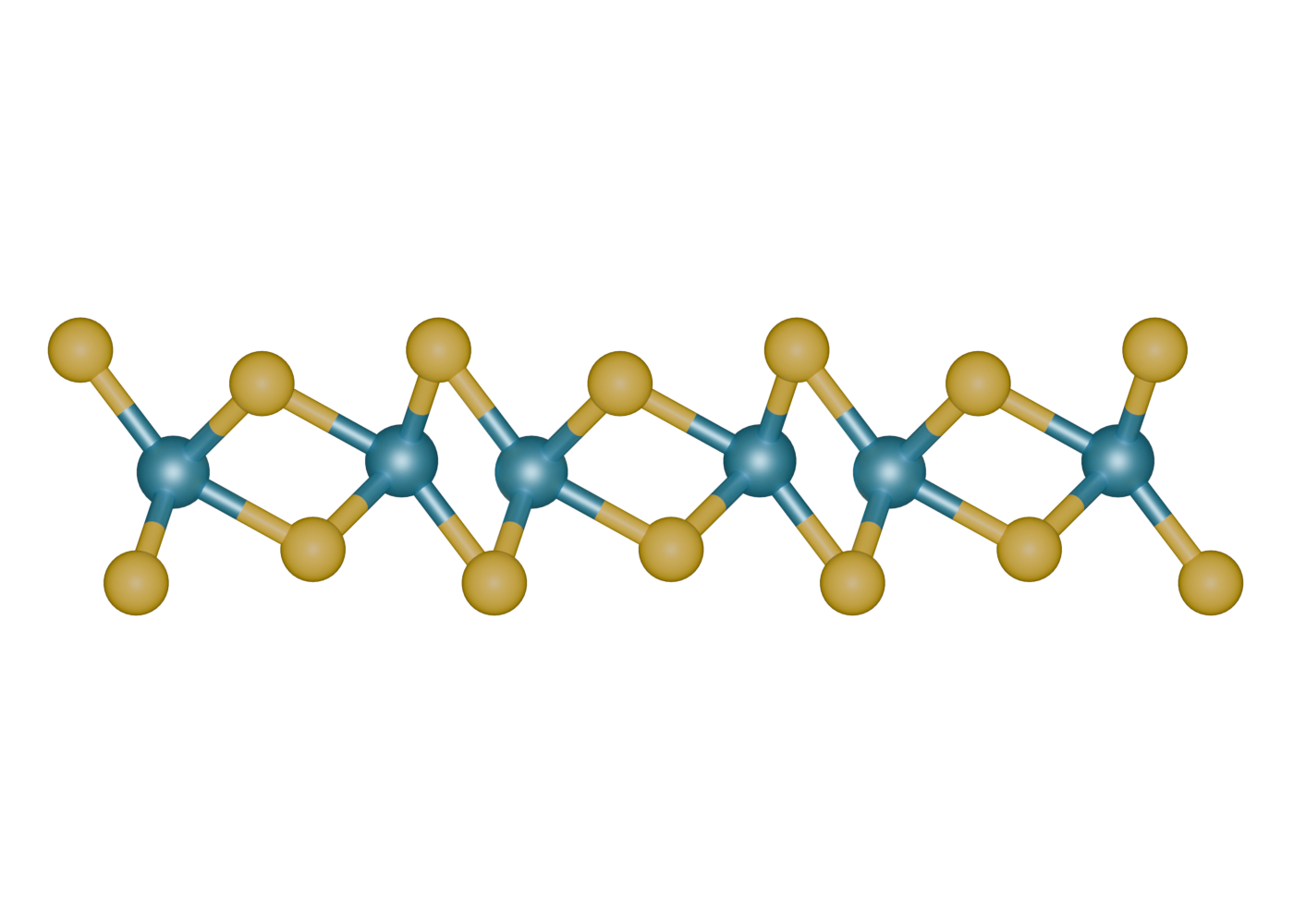}\\ \hline
     Odd \# ML & $\chi^2 \neq 0$ & $\chi^2 \neq 0$ & $\chi^2 = 0$ \\ \hline
     Even \# ML & $\chi^2 = 0$ & $\chi^2 \neq 0$ & $\chi^2 \neq 0$ \\ \hline
     \centeredtxta{Nonlinear($\phi$)}& \includegraphics[width=0.25\linewidth]{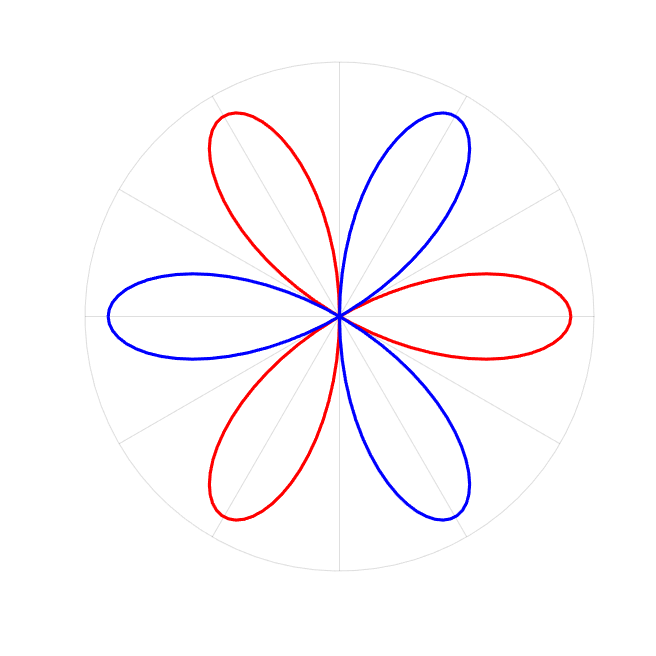}& \includegraphics[width=0.25\linewidth]{figs/sim_3f.png} & \includegraphics[width=0.25\linewidth]{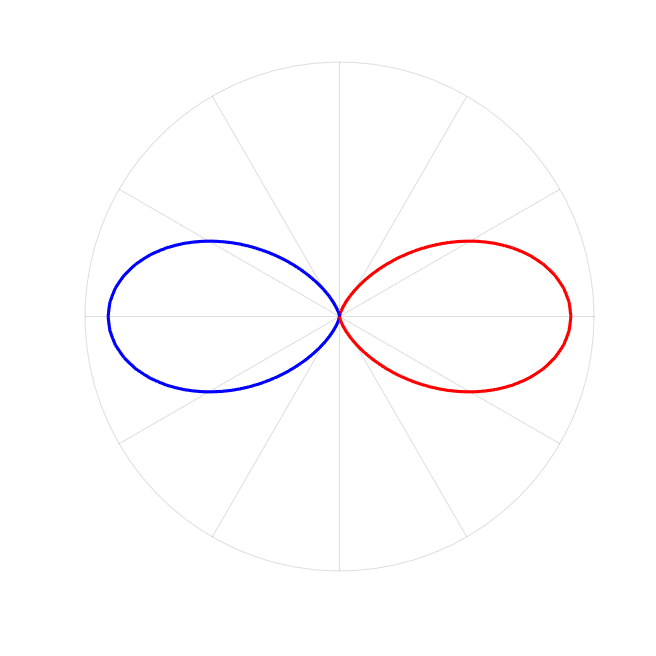}\\ \hline \hline
     \centeredtxta{Magnetic($\phi$) (TI/TMD   /FM)}& \includegraphics[width=0.25\linewidth]{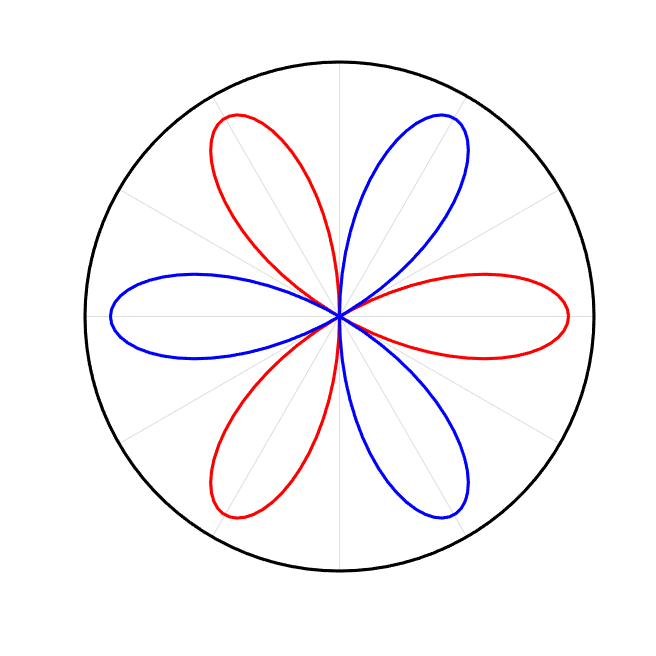} & \includegraphics[width=0.25\linewidth]{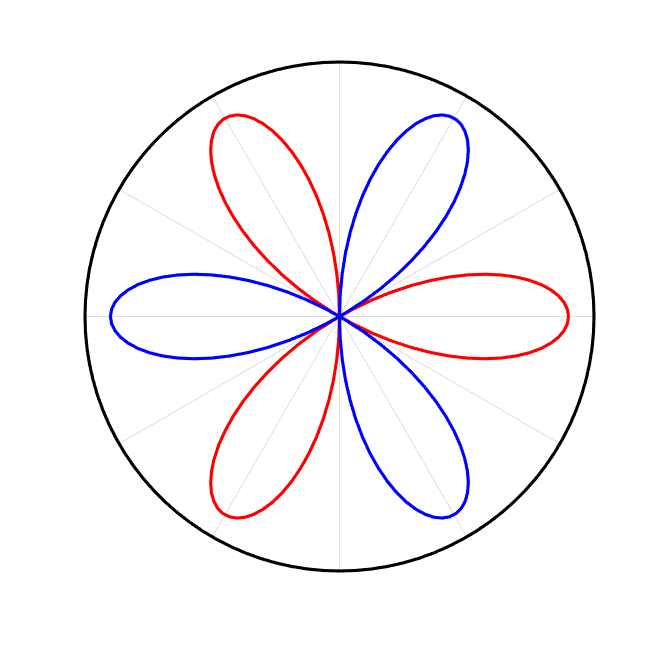} & 
     \includegraphics[width=0.25\linewidth]{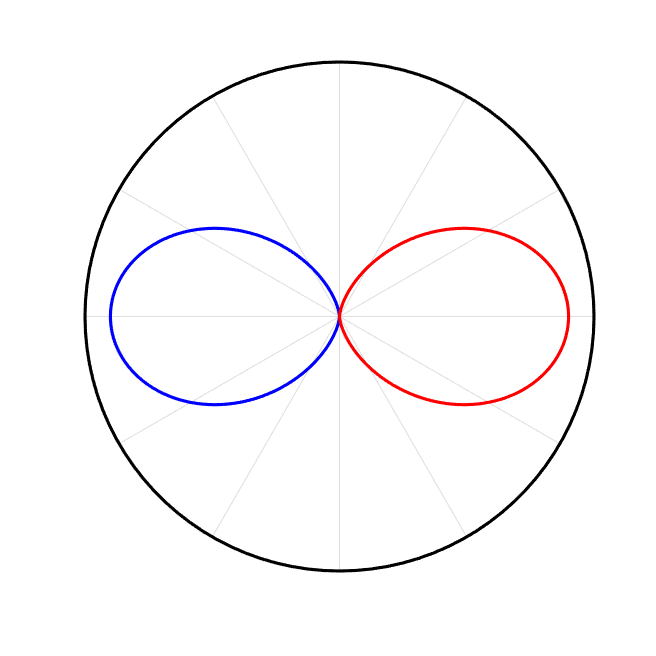}

\end{tabular}
\caption{\textbf{Comparison of different polytypes 2H, 3R and 1T$^\prime$ of \WSe2}. The first and second row show the struture of the layers (top and side views), the third and fourth show the behaviour of second order nonlinearity $\chi^2$ for different number of layers, the fifth row shows the  calculated azimuthal dependence of the real part of the nonlinearity (red = positive, blue = negative values). The sixth row shows the expected magnetic response of the stack with the azimuthal dependence of the SCC (black) and magnetic dependent nonlinearity (red/blue curve).
}
    \label{tab_layer}
\end{table}

\section{\label{sec:Mat} Growth and structural characterization of epitaxial TI/TMD/FM heterostructures.}

All the \BISE(10QL)/\WSe{2}($t_\text{TMD}$) heterostructures were grown by molecular beam epitaxy (MBE) on large area c-cut sapphire substrates, with the details provided in the Methods section (QL - quintuple layers). These substrate types have shown the possibility of oriented TMD epitaxial growth for wafer scale single crystal films owing to their miscut nature~\cite{Wan22_wafer}. $t_\text{TMD}$ corresponds to the \WSe{2} layer thickness (1, 2, 3 and 4 MLs). A low growth temperature was chosen to favor the presence of multi-polymorphs of \WSe{2}. 3~nm of cobalt (Co) and aluminum (Al) were then evaporated by e-beam at room temperature at 0.1 {\AA}/s  and 0.5 {\AA}/s rates, respectively. The aluminium oxidises to provide a AlO$_x$ protective insulating layer of the heterostructure.

\begin{figure}[!h]
  \begin{center}
      \includegraphics[width=\textwidth]{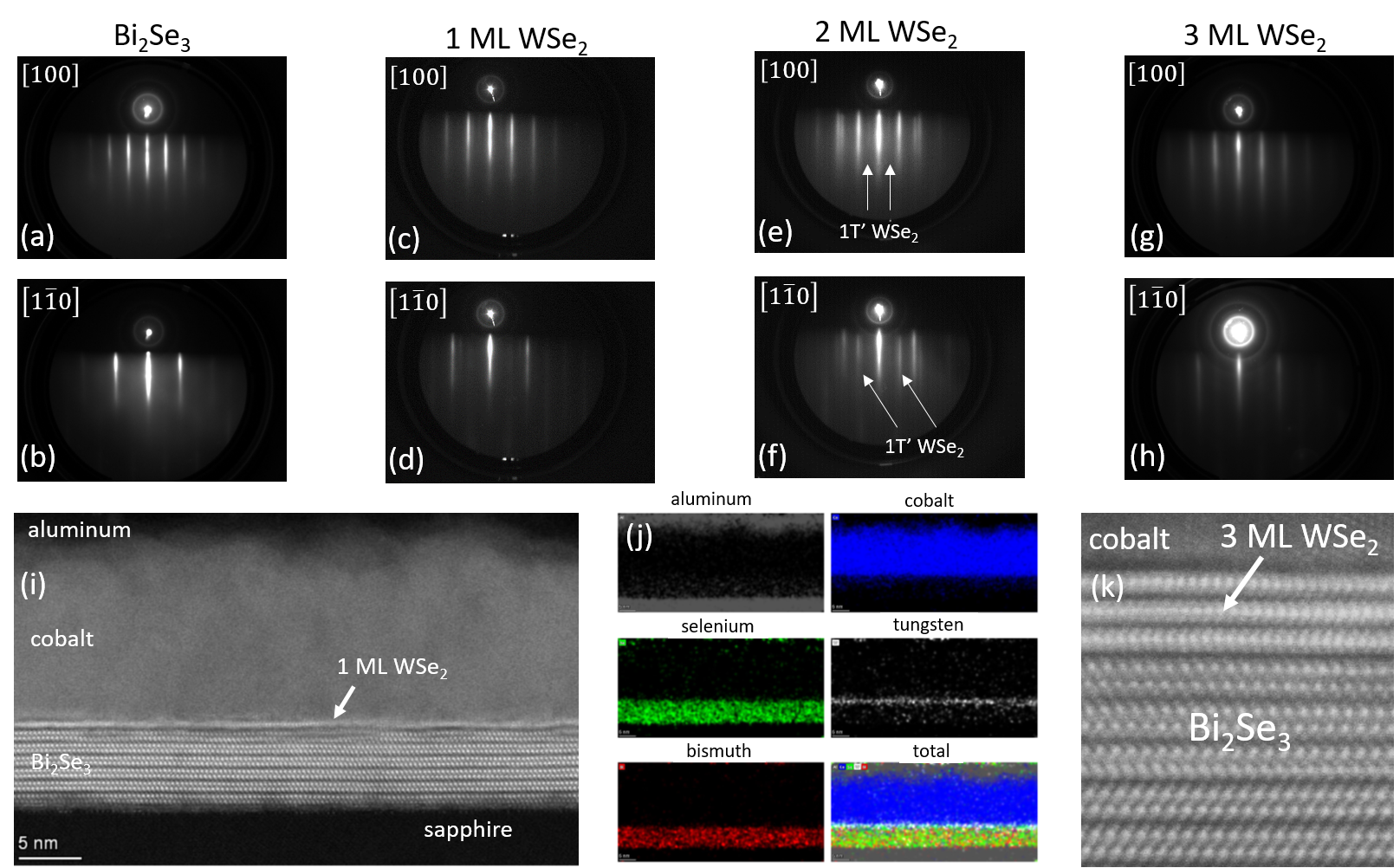}
       \caption{\textbf{Structure and polymorph characterisation of TI/TMD/FM heterostructure}. (a-h) RHEED patterns recorded along two azimuths [100] and [1$\bar{1}$0] of 10 QL of \BISE~and 1-3 ML of \WSe{2}. (i) STEM cross section image of \BISE(7QL)/\WSe{2}(1ML)/Co(10nm)/Al(3nm). (This stack had a thinner \BISE~and a thicker Co layer than that tested for the THz measurements). (j) Corresponding elemental chemical analysis by EDX of the \BISE(7QL)/\WSe{2}(1ML)/Co(10nm)/Al(3nm) stack: aluminum (dark grey), cobalt (blue), selenium (green), tungsten (white) and bismuth (red). (k) Atomic scale STEM cross section image of \BISE(10QL)/\WSe{2}(3ML)/Co(3nm)/Al(3nm) showing the vdW gaps, Co/\WSe{2} and \WSe{2}/\BISE~interfaces. }
    \label{fig_croissance_RHEED}
  \end{center}
\end{figure}

The reflection high energy electron diffraction (RHEED) patterns are shown in Figure~\ref{fig_croissance_RHEED}(a-h). The two distinct patterns along [100] and [1$\bar{1}$0] azimuths demonstrate the single crystalline character of the layers. We also clearly see the epitaxial relationship between the successive layers and the presence of thin streaks indicates the atomically flat character of the films as expected for the layer-by-layer growth of 2D materials. Co and Al are polycrystalline with RHEED patterns exhibiting rings . Regarding the different possible polymorphs of WSe$_2$, in the RHEED patterns recorded along the [100] direction, we observe faint lines in-between the first order diffraction lines resembling a (x2) surface reconstruction. This diffraction pattern is characteristic of the presence of the 1T$^\prime$ polymorph of \WSe{2} at the surface of the film~\cite{tang_quantum_2017,chen_1Tgrowth_2019,chen_large_2018}. The low growth temperature promotes the formation of this polymorph and is also necessary to avoid damaging the \BISE~layer that starts decomposing above 300°C. The presence of the 1T$^\prime$ polymorph is more visible for 2 ML of \WSe{2} as highlighted in Figure~\ref{fig_croissance_RHEED}(e,f). Scanning transmission electron microscopy (STEM) images in cross section are shown in Figure~\ref{fig_croissance_RHEED}(i) and (k) for 1 ML and 3 ML of \WSe{2}, respectively. We can clearly observe van der Waals gaps between the \BISE~QLs and \WSe{2} MLs. Moreover, as shown in Figure~\ref{fig_croissance_RHEED}(k), Co/\WSe{2} and \WSe{2}/\BISE~interfaces are atomically sharp and appear as quasi-vdW gaps in the STEM image. Regarding other stacking orders, the 2H polymorph is observed for 2 MLs and a mixture of 3R and 2H polymorphs for layers greater than 2 MLs. The details about the stacking of \WSe{2} layers are given in the Supp. Info. In Figure~\ref{fig_croissance_RHEED}(j), the elemental chemical analysis by energy dispersive x-ray spectroscopy (EDX) corresponding to the STEM image of Figure~\ref{fig_croissance_RHEED}(i) confirms the absence of atomic intermixing at the Co/\WSe{2} interface and the role played by \WSe{2} as a diffusion barrier between the ferromagnet and the TI avoiding selenium diffusion into the cobalt layer.

To summarise, 1T$^\prime$ is present in all samples, 2 MLs has a 2H polymorph and thicker layers shows a mixture of 3R and 2H polymorphs. Raman spectroscopy was also performed to verify the monolayer nature and thickness of the deposited WSe$_2$ spacer (see Supp. Info.). The evolution of the Raman shifts are in close agreement with the literature on WSe$_2$ layers~\cite{zhao_lattice_2013}.

\section{\label{sec:THz} Coherent current transport via THz emission spectroscopy}

We now move to down-conversion in these Bi$_2$Se$_3$/WSe$_2$/Co heterostructures through coherent THz emission spectroscopy where an ultrafast optical excitation is downconverted to the THz range via ultrafast currents through two particular phenomena:

\textit{i}) Optical nonlinear conversion where a resonant or non-resonant optical excitation results in second order nonlinearities for ultrafast current generation through optical rectification for THz generation, similar to SHG:
\begin{equation}
    E_{\text{THz}} \propto \chi^{(2)}_{total}(\phi) E_{NIR} E_{NIR}^*
\end{equation}
where $E_{\text{THz}}$ is the emitted THz field and $E_{NIR}$ corresponds to the electric field of the optical excitation, and

\textit{ii}) Spintronic THz emission that relies on the ultrafast demagnetization of the thin ferromagnetic layer and transient SCC in a strong spin-orbit coupling material via
\begin{equation}
    E_{\text{THz}} \propto \theta_{\text{SCC}}\left(j_s \times \frac{\mathbf{M}}{\abs{\mathbf{M}}}\right)
\end{equation}
where $j_s$ is the spin current, $\mathbf{M}$ is the magnetization vector and $\theta_{\text{SCC}}$ is the spin-Hall angle for structures where the spin-Hall effect occurs~\cite{seifert_efficient_2016, dang_ultrafast_2020}, such as in heavy metals. The latter is replaced by the inverse Rashba-Edelstein length ($\lambda_{\text{IREE}}$) for SCC by the inverse Rashba-Edelstein effect (IREE) in, for example, the surface states of TIs~\cite{Zhou2018,wang_ultrafast_2018,tong_enhanced_2020,chen_efficient_2021,rongione_ultrafast_2022}. 

Note that coherent THz emission spectroscopy measures directly the amplitude and phase of the generated pulses (unlike SHG) and has become a powerful non-contact technique to probe the current and spin-injection properties in complex spintronic heterostructures~\cite{dang_ultrafast_2020}, including 2D structures~\cite{cheng_far_2019} to the examples above. In the studied heterostructures, the THz spintronic emision occurs owing to SCC at the TI interface and the TMD layer, and the THz nonlinear emission is a result of both the \BISE~and WSe$_2$ layers. 
The THz emission of our samples was measured on ytterbium based OPA system (780~nm) with ZnTe electro-optic detection. An electromagnet was used to align domains in ferromagnetic layer (see Supp. Info. for more information). All measurements were performed at room temperature.

\begin{figure}[!h]
\centering
  \begin{subfigure}{1\linewidth}
  \centering
      \includegraphics[width=1\linewidth]{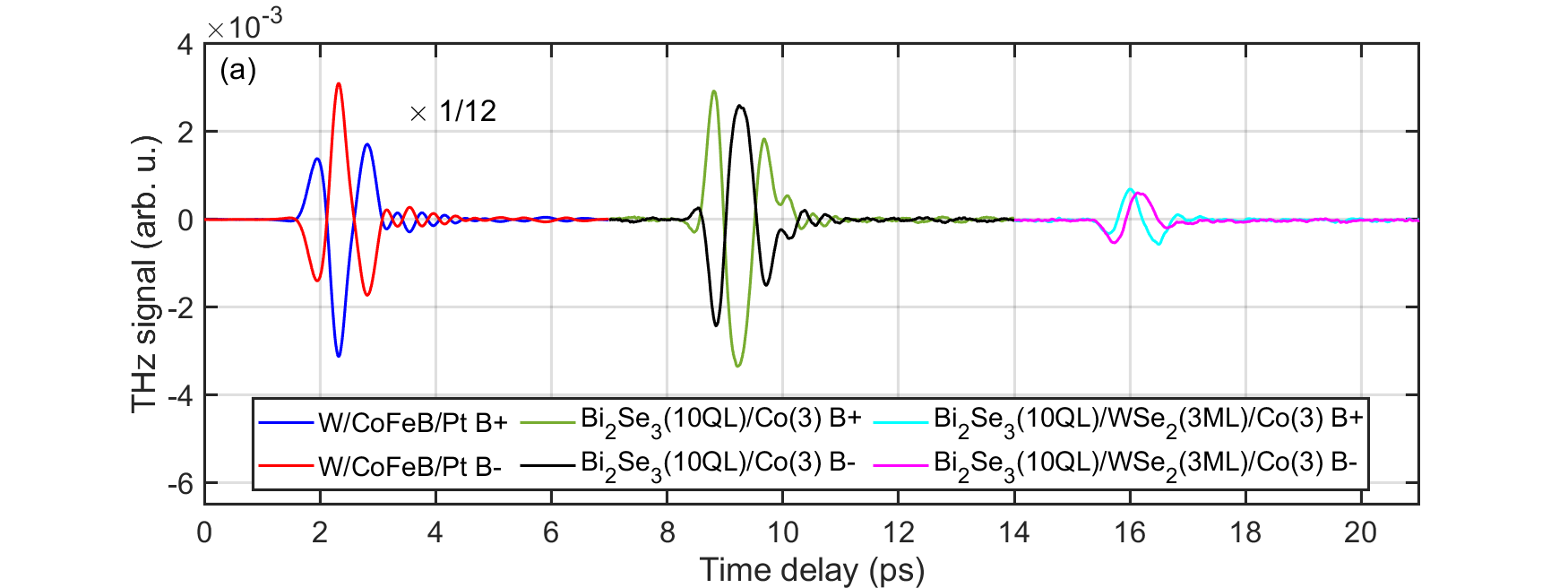}
  \end{subfigure}
  \begin{subfigure}{0.49\linewidth}
  \centering
      \includegraphics[width=1\linewidth]{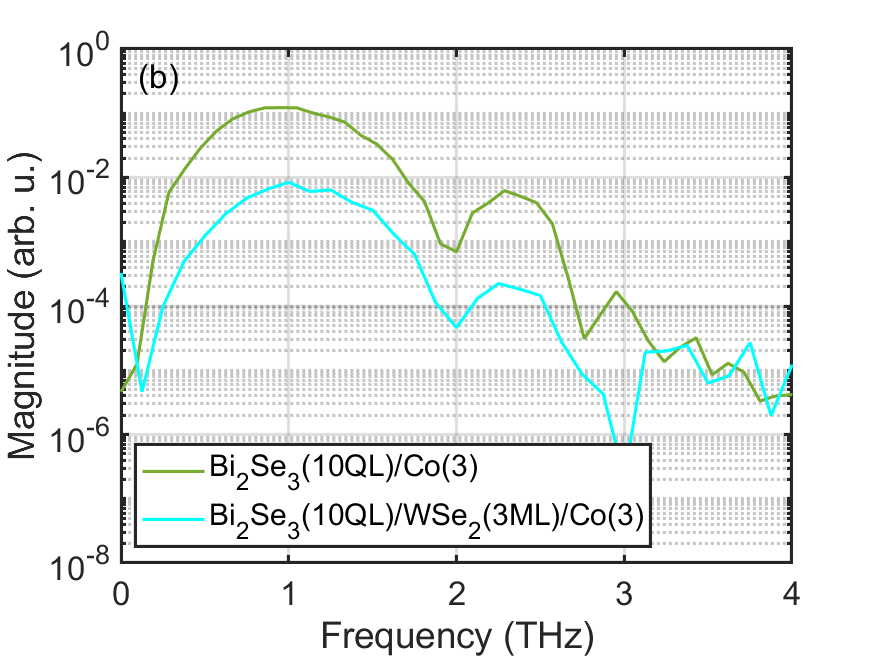}
  \end{subfigure}
  \begin{subfigure}{0.49\linewidth}
  \centering
      \includegraphics[width=1\linewidth]{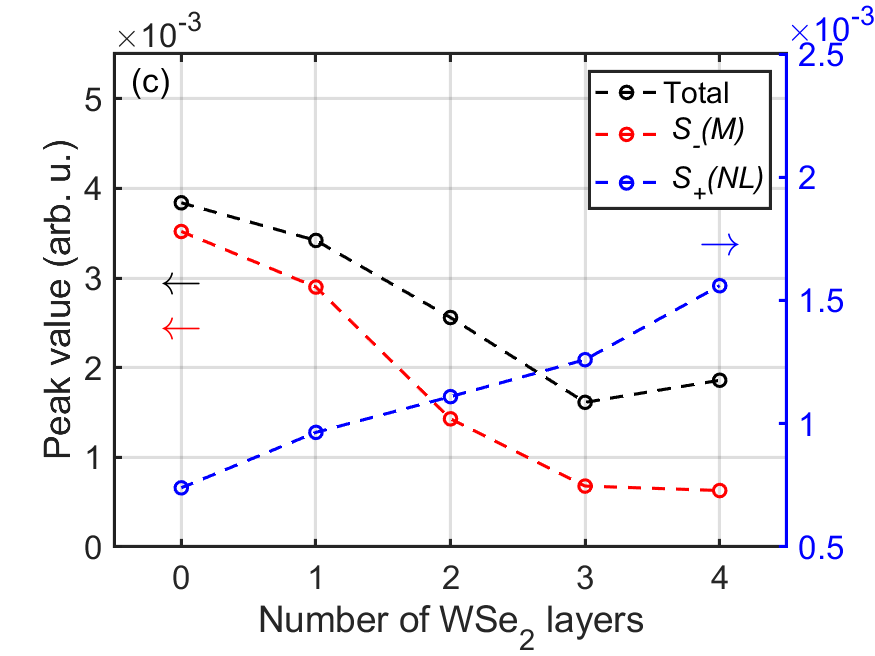}
  \end{subfigure}
       \caption{\textbf{Evolution of THz amplitude for nonlinear and spintronic processes with WSe$_2$ barrier thickness}. (a) THz emission from \BISE/Co/AlO$_x$ and \BISE/\WSe{2}(3ML)/Co/AlO$_x$ compared to W/CoFeB/Pt at two opposite magnetic field polarities B$\pm$. The curves for each sample are shifted in time for clarity. (b) THz spectrum from the \BISE/Co/AlO$_x$ and \BISE/\WSe{2}(3ML)/Co/AlO$_x$ samples showing a bandwidth of approximately 3~THz (detection limited). (c) Evolution of the total (black), $S_{-}(M)$ (red) and $S_{+}(NL)$ (blue curve) contributions from \BISE/\WSe{2}($t_\text{TMD}$)/Co/AlO$_x$ as a function of the \WSe{2} number of layers, $t_\text{TMD}$, ranging from 0 to 4 ML. All stacks had 10 QLs of \BISE, 3nm of Co and 3nm of AlO$_x$ }
    \label{fig_THz_tab}
\end{figure}

We first present in Figure~\ref{fig_THz_tab}(a) the THz electric field emission from Bi$_2$Se$_3$/Co and Bi$_2$Se$_3$/\WSe{2}/Co compared to an optimised metallic spintronic structure of W/CoFeB/Pt (2/1.8/2~nm thick) to highlight the behaviour of nonlinear and magnetic response. Two set of curves for each sample is shown, corresponding to flipping of the applied magnetic field direction. For the metallic structure a change in sign of the THz phase is seen between opposite magnetic directions, indicating a spin-charge conversion process and is entirely of magnetic origin. 
 
Compared to W/CoFeB/Pt, the THz phase obtained from Bi$_2$Se$_3$/Co is \textit{i}) the same for a given magnetic field polarity, demonstrating an interconversion sign identical to the spin Hall effect sign of Pt, and \textit{ii}) does not fully reverse with the magnetic field, i.e. the absolute value of the E-field amplitudes are not identical, which indicates the influence of the additional nonlinear optical contributions to the measured THz signal. The THz emitted signal from Bi$_2$Se$_3$/\WSe{2}(3ML)/Co shows the opposite extreme where no sign change in phase is observed and the signals for opposite applied magnetic fields are almost equal, indicating a small magnetic and large nonlinear component. The typical spectra for vdW structures are shown in the Figure~\ref{fig_THz_tab}(b), showing a bandwidth of $\sim$ 3~THz limited by the optical excitation and detector. As discussed below, the nonlinear behaviour is enhanced in the heterostructures with \WSe{2}.

The nonlinear $S_+(NL)$ and magnetic $S_-(M)$ contributions to the time domain signal can be extracted~\cite{rongione_ultrafast_2022} through the sum and difference, respectively, of the measured phase resolved THz signals at +B and -B applied magnetic field. (By changing the polarity of the magnetic field, the magnetization of Co follows the applied field direction owing to its soft magnetic character) i.e.:
\begin{equation}
    S_+(NL) = \frac{S_\text{THz}(+B)+S_\text{THz}(-B)}{2} \quad \text{and} \quad S_-(M) = \frac{S_\text{THz}(+B)-S_\text{THz}(-B)}{2}.
    \label{eq:components}
\end{equation}

\begin{figure}[h]
    \centering
    \begin{subfigure}{0.49\linewidth}
    \centering
        \includegraphics[width=\linewidth]{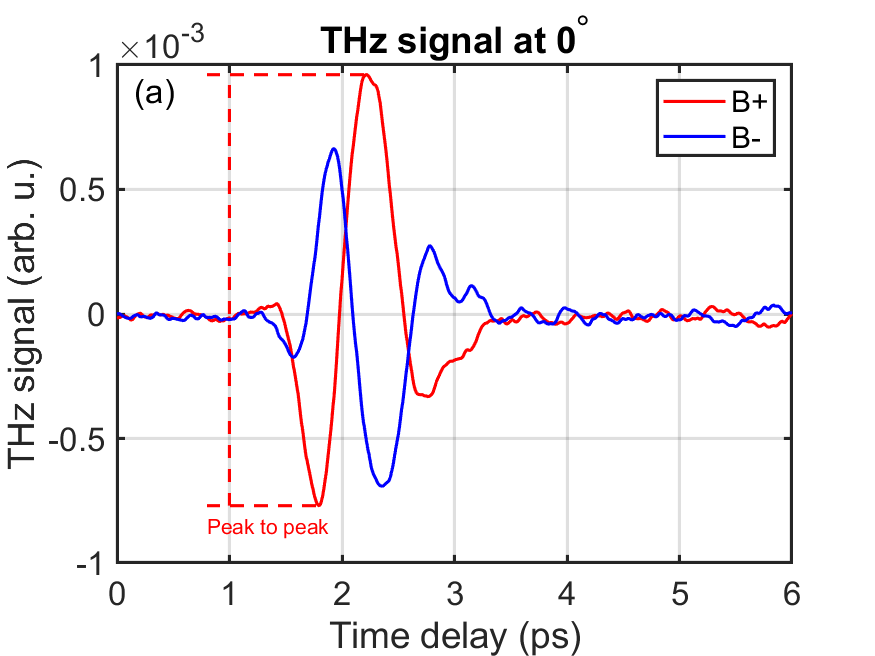}
    \end{subfigure}
    \begin{subfigure}{0.49\linewidth}
    \centering
        \includegraphics[width=\linewidth]{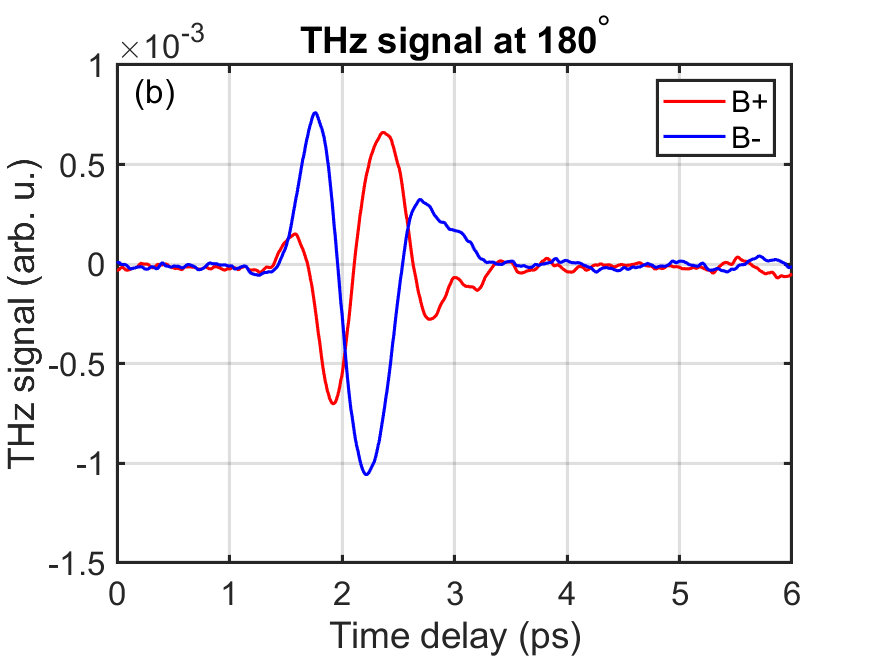}
    \end{subfigure}
    \begin{subfigure}{0.49\linewidth}
    \centering
        \includegraphics[width=\linewidth]{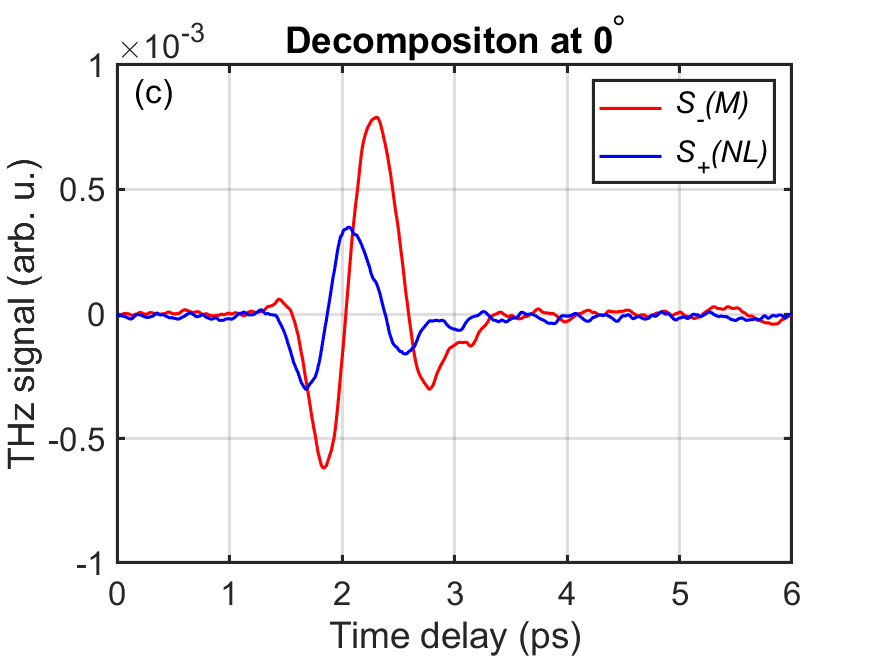}
    \end{subfigure}
    \begin{subfigure}{0.49\linewidth}
    \centering
        \includegraphics[width=\linewidth]{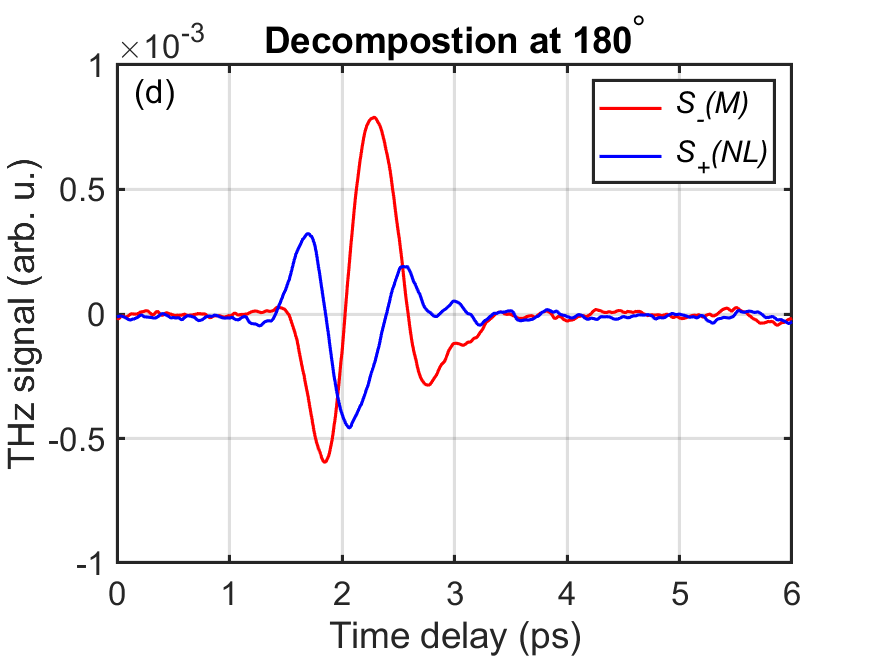}
    \end{subfigure}

    \caption{\textbf{Example of extraction of magnetic $S_-(M)$ and nonlinear $S_+(NL)$} (a,b) THz signal for sample \BISE/\WSe{2}(2ML)/Co sample at two azimuth angles (0$^\circ$ and 180$^\circ$). (c,d) Extracted and decomposed into $S_-(M)$ and $S_+(NL)$ signals at the two angles using Eq.~\ref{eq:components}.}
    
    \label{fig_THz_contri}
\end{figure}

An example of the typical extraction of the nonlinear and magnetic components is shown in Figure~\ref{fig_THz_contri} for Bi$_2$Se$_3$/\WSe{2}(2ML)/Co sample. Figure~\ref{fig_THz_contri}(a,b) shows the acquired time domain signal for two different azimuthal angles, and (c,d) shows the extracted magnetic contribution and the nonlinear contribution for each angle. Note the difference in phase of nonlinear emission between these particular azimuthal angles, whilst the THz emission of magnetic origin remains almost identical. From this extraction, Figure~\ref{fig_THz_tab}(c) shows peak-to-peak nonlinear and magnetic contributions as a function of the WSe$_2$ inset thickness, ranging from 0 ML to 4 ML, for their maximum measured values with sample azimuthal rotation (see below). The figure shows that the magnetic component $S_{-}(M)$ drops rapidly with \WSe{2} layers as it acts as a spin barrier, while the nonlinear component $S_{+}(NL)$ increases with number of layers as each layer contributes to the nonlinear signal.

\begin{figure}[h]
    \centering

    \begin{subfigure}{0.32\linewidth}
        \centering
        \includegraphics[width=1\linewidth]{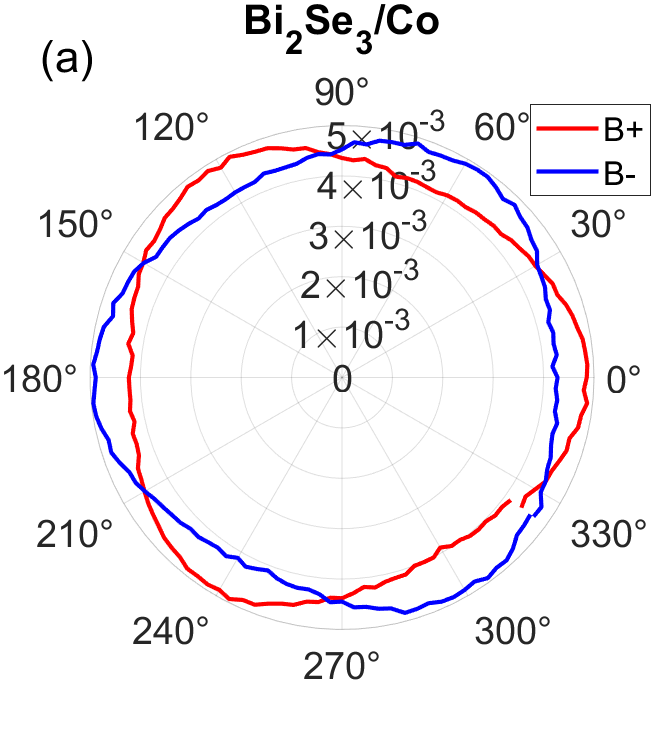}
    \end{subfigure}
    \begin{subfigure}{0.32\linewidth}
        \centering
        \includegraphics[width=1\linewidth]{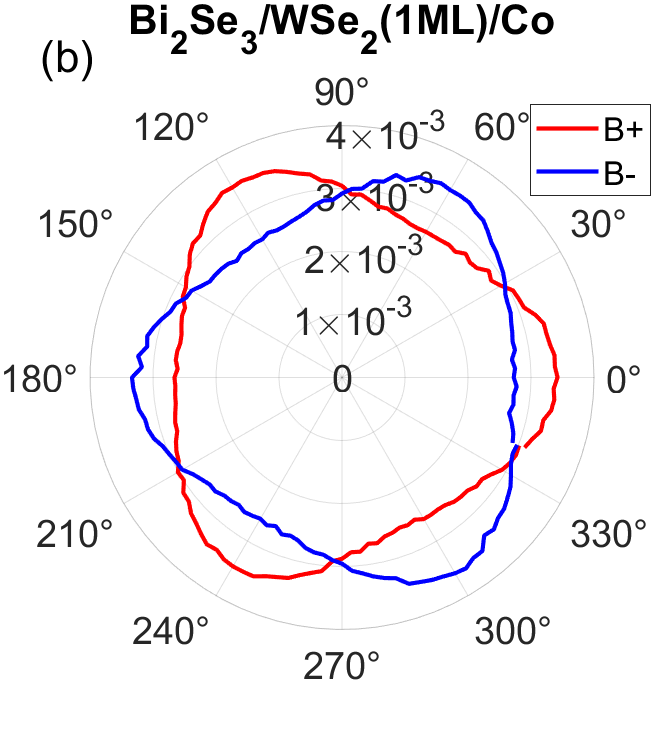}
    \end{subfigure}
    \begin{subfigure}{0.32\linewidth}
        \centering
        \includegraphics[width=1\linewidth]{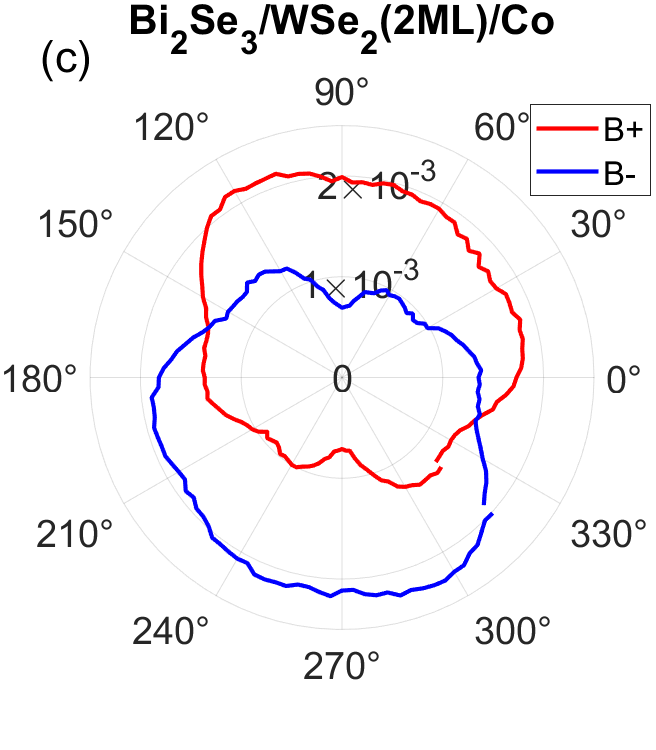}
    \end{subfigure}
    \begin{subfigure}{0.32\linewidth}
        \centering
        \includegraphics[width=1\linewidth]{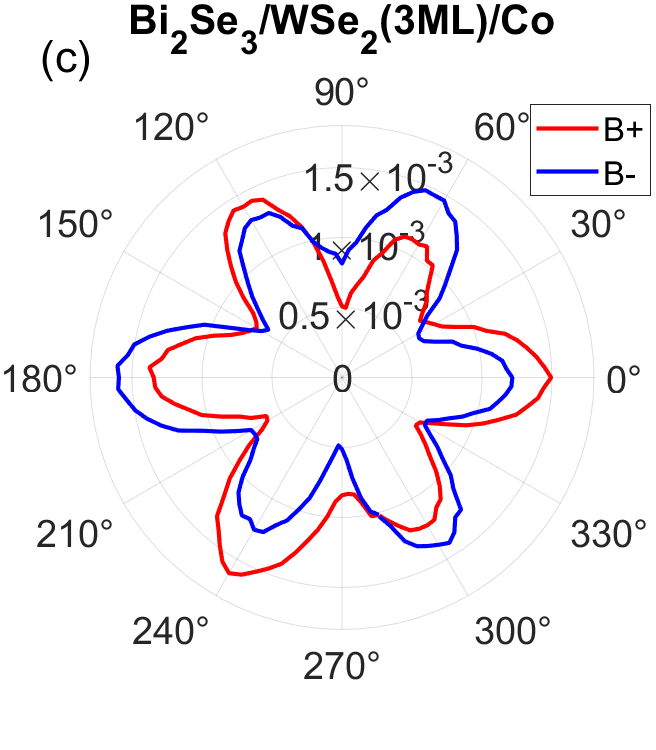}
    \end{subfigure}
    \begin{subfigure}{0.32\linewidth}
        \centering
        \includegraphics[width=1\linewidth]{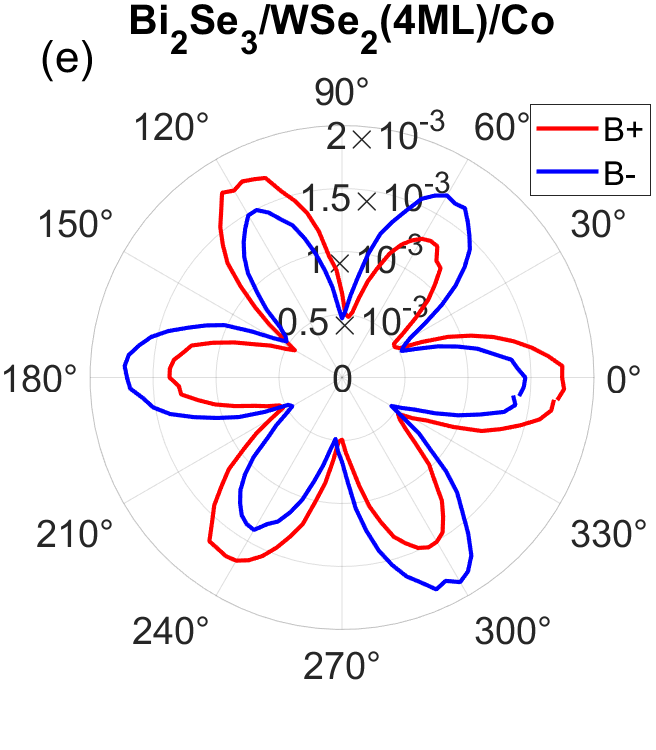}
    \end{subfigure}
    
    \caption{\textbf{Peak-to-peak total amplitude} in absolute values of emitted THz pulses from \BISE/WSe$_2$/Co heterostructures for thicknesses of WSe$_2$ ranging from 0 ML to 4 MLs (a-e) as a function of azimuthal angle rotation for two opposite orientations of the external magnetic field B$\pm$. The azimuth angle of the samples is oriented arbitrary, with positive lobe as origin.}
    \label{fig:total}
\end{figure}

Beyond absolute values of the THz field, the effect of the \WSe{2} polymorph on the emission symmetries of both magnetic $S_-(M)$ and nonlinear $S_+(NL)$ components were investigated by measuring the THz peak-to-peak field as a function of the crystallographic azimuthal angle $\phi$. The results are reported in Figure~\ref{fig:total} showing the peak-to-peak THz field values for two opposite magnetic field directions. Large changes are seen in the symmetries when transitioning from 0 ML to 4 MLs of inserted \WSe{2}. To decouple the magnetic and nonlinear components of the signals, we followed the same procedure as above (Figure~\ref{fig_THz_contri} and Equation~\ref{eq:components}) to compare the THz signals with opposite applied magnetic fields with the results shown below. 

\begin{figure}[h]
    \centering
    \begin{subfigure}{0.32\linewidth}
        \centering
        \includegraphics[width=1\linewidth]{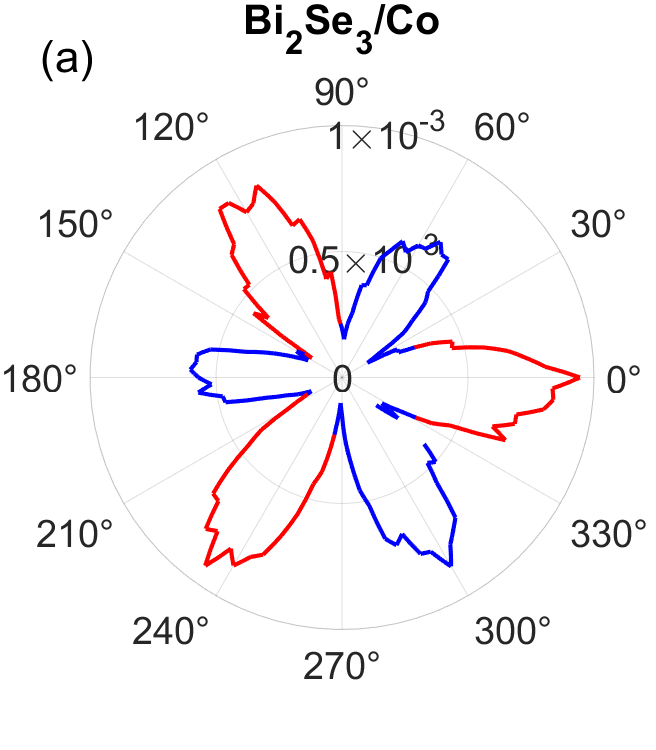}
    \end{subfigure}
    \begin{subfigure}{0.32\linewidth}
        \centering
        \includegraphics[width=1\linewidth]{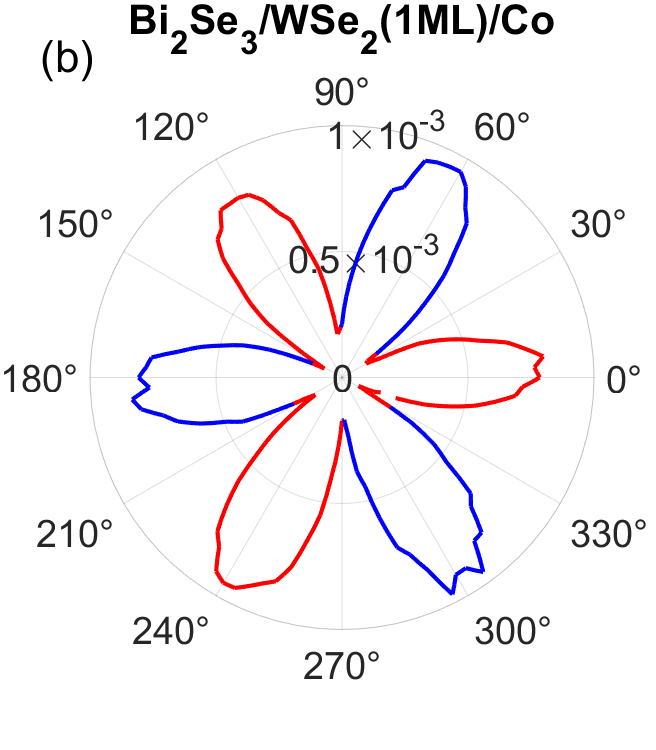}
    \end{subfigure}
    \begin{subfigure}{0.32\linewidth}
        \centering
        \includegraphics[width=1\linewidth]{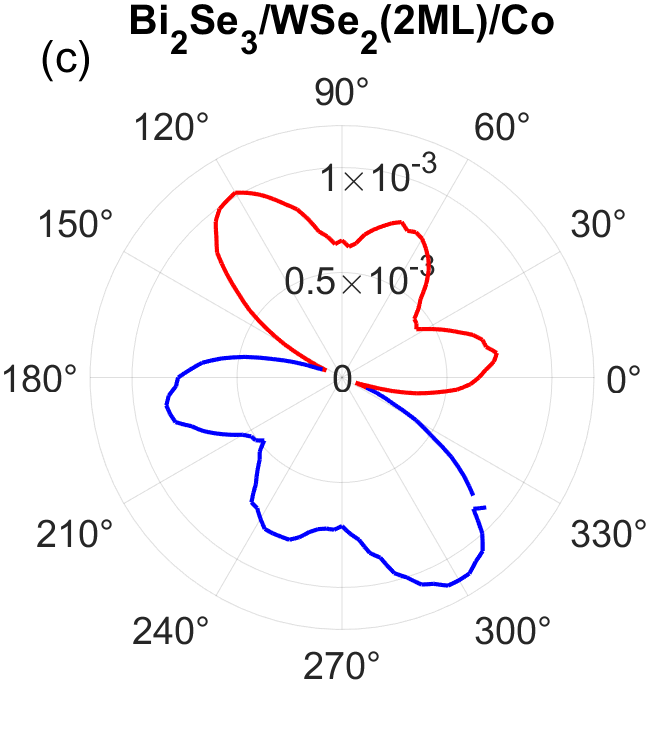}
    \end{subfigure}
    \begin{subfigure}{0.32\linewidth}
        \centering
        \includegraphics[width=1\linewidth]{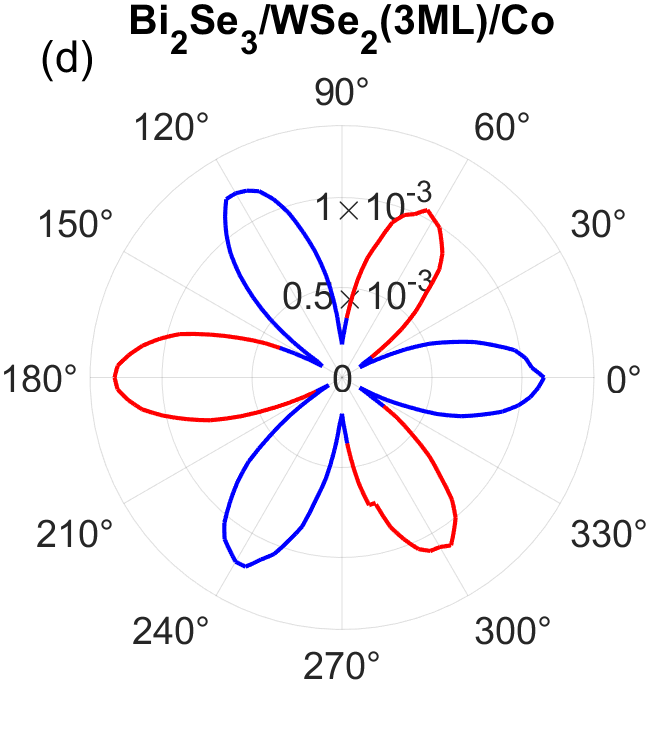}
    \end{subfigure}
    \begin{subfigure}{0.32\linewidth}
        \centering
        \includegraphics[width=1\linewidth]{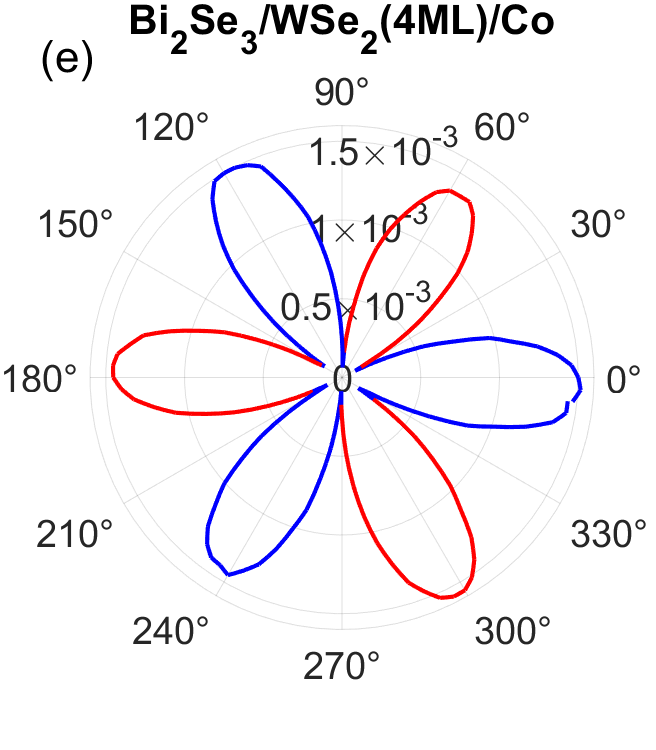}
    \end{subfigure}
    \caption{\textbf{Peak to peak amplitudes of the nonlinear components} of the THz signal $S_+(NL)$ decomposed by Eq.~\ref{eq:components}), for samples with different number of \WSe2 layers, (a-e) corresponding to 0 - 4 ML. The color of lines represents positive (red) and negative (blue) polarity (see Supp. Info.).}
    \label{fig:nonmag}
\end{figure}

\begin{figure}[h]
    \begin{subfigure}{0.32\linewidth}
        \centering
        \includegraphics[width=1\linewidth]{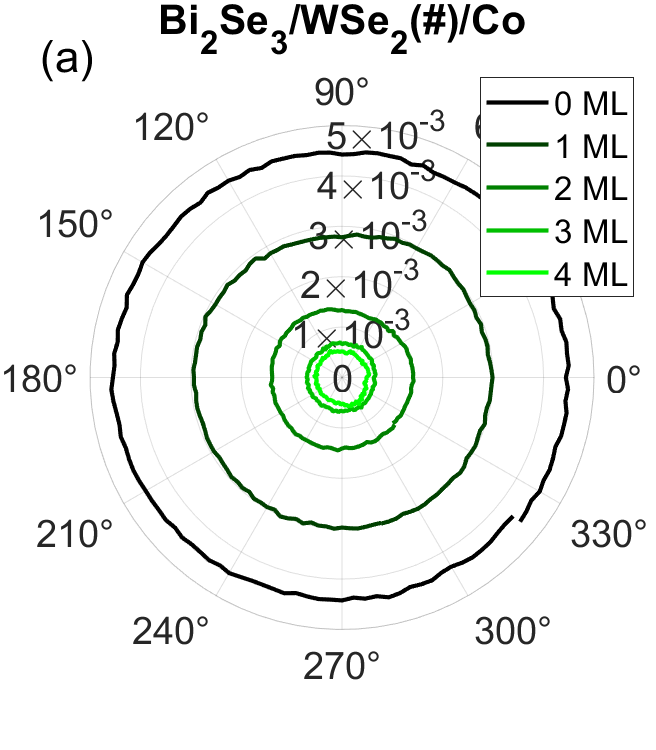}
    \end{subfigure}
    \begin{subfigure}{0.32\linewidth}
        \centering
        \includegraphics[width=1\linewidth]{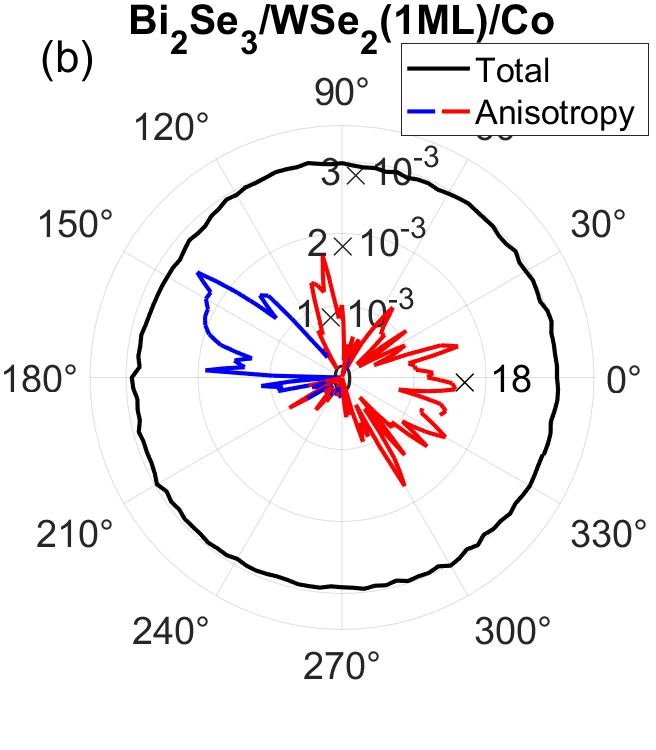}
    \end{subfigure}
    \begin{subfigure}{0.32\linewidth}
        \centering
        \includegraphics[width=1\linewidth]{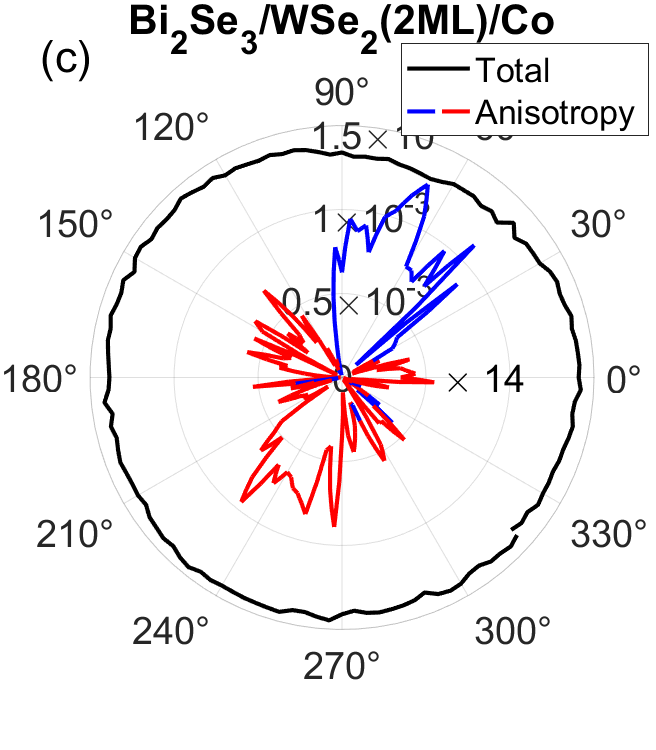}
    \end{subfigure}
    \caption{\textbf{Peak to peak amplitudes of the magnetic components} of the THz signal $S_-(M)$ decomposed by Eq.~\ref{eq:components}) for (a) all samples with different number of \WSe2 layers. (b,c) Comparison of anisotropic component of magnetic signal for 1 and 2 ML obtained by removing the isotropic part (i.e. average value of black curve).}
    \label{fig:mag}
\end{figure}

First, on the nonlinear contributions $S_{+}(NL)$ (Figure~\ref{fig:nonmag}), we observe that on Bi$_2$Se$_3$/Co without the WSe$_2$ layer, the symmetries of the nonlinear contributions $S_{+}(NL)$ are six-fold with a small asymmetry between adjacent lobes, in agreement with previous studies of SHG from \BISE~whose origin is from the surface states~\cite{Hsieh11_nonlinear_TI}. The red and blue points correspond to positive and negative values of the THz field respectively (see Supp. Info). The magnetic component is isotropic with rotation, as expected for spintronic emission (see Figure~\ref{fig:mag}(a)). When a WSe$_2$ monolayer spacer is inserted in Figure~\ref{fig:nonmag}, the nonlinear contribution is changed considerably owing to the WSe$_2$ high second-order electric susceptibility $\chi^{(2)} \simeq 5$ nm.V$^{-1}$ determined through SHG ~\cite{ribeiro-soares_second_2015}. For increasing $t_\text{TMD}$ we observe that the nonlinear contributions increases compared to Bi$_2$Se$_3$/Co, presenting a six-fold symmetry for 1, 3 and 4 ML. This corresponds to the expected three-fold symmetry for 2H and 3R polymorphs of \WSe{2}, as observed for SHG, noting that up to 2 MLs, \WSe{2} shows a 2H form and increasing thicknesses show a mix of 2H and 3R polymorphs. As discussed above, space-inversion symmetry as a function of the parity of the WSe$_2$ ML thickness plays an important role. Indeed, for the 2H polymorph, an even number of MLs presents space-inversion symmetry (point group $D_{3d}$) and an odd number of MLs, WSe$_2$ breaks space-inversion symmetry (point group $D_{3h}$), whilst for the 3R case, space-inversion symmetry is always broken, as reported previously in Refs.~\cite{ribeiro-soares_second_2015,rosa_characterization_2018}. Therefore, a six fold optical emission response based on optical nonlinearities is expected for any 3R form and only for odd number of layers for the 2H polymorph. However, for WSe$_2$(2ML), not only a strong second order response is observed but the symmetry drastically changes to two-fold, which would be incompatible with nonlinearities for 2H geometries and for an even number of \WSe{2} layers. This second order effect is therefore solely owing to the 1T$^\prime$ polymorph observed in these layers that is expected to give a two fold symmetry (see Table~\ref{tab_layer}). In 1T$^\prime$ form, the trend of layer dependent SHG is opposite to that in 2H phase i.e. even layers of 1T$^\prime$ with inversion symmetry breaking display significant SHG while odd layers with inversion symmetry show negligible SHG~\cite{ribeiro-soares_second_2015,song_extraordinary_2018}. Note that 4 MLs \WSe{2} sample shows a three fold symmetry as the 3R polymorph is dominant and where the 2H polymorph is expected to give a zero nonlinearity owing to the symmetry. Therefore, this works shows how the THz nonlinearities in these junctions can be controlled at the monolayer level through the structural polymorphs of \WSe{2} and stacking order. We emphasize that the thickness-dependent THz emission (down-conversion) is reported here for the first time where all previous works have been on SHG, showing the high crystalline quality of the material highlighted by the fact that the nonlinearities are present over large areas (THz spot size $\sim$ 1000~\textmu m). This shows also that THz emission spectroscopy and its inherent polarization sensitivity can be an powerful probe to determine the thickness parity and symmetries of 2D materials on macroscopic scales.

\begin{figure}[h]
    \centering
    \begin{subfigure}{0.49\linewidth}
    \centering
        \includegraphics[width=\linewidth]{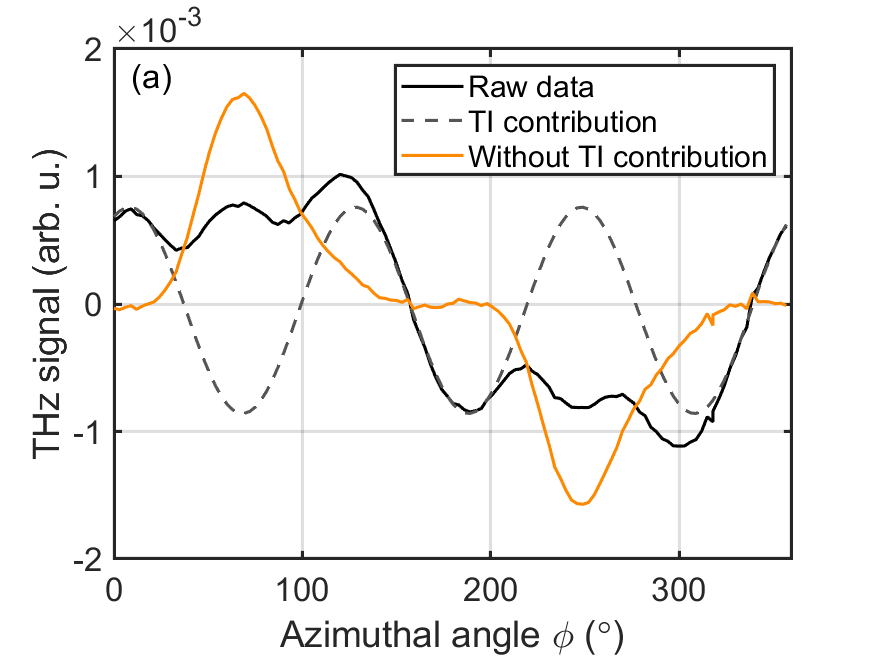}
    \end{subfigure}
    \begin{subfigure}{0.49\linewidth}
    \centering
        \includegraphics[width=\linewidth]{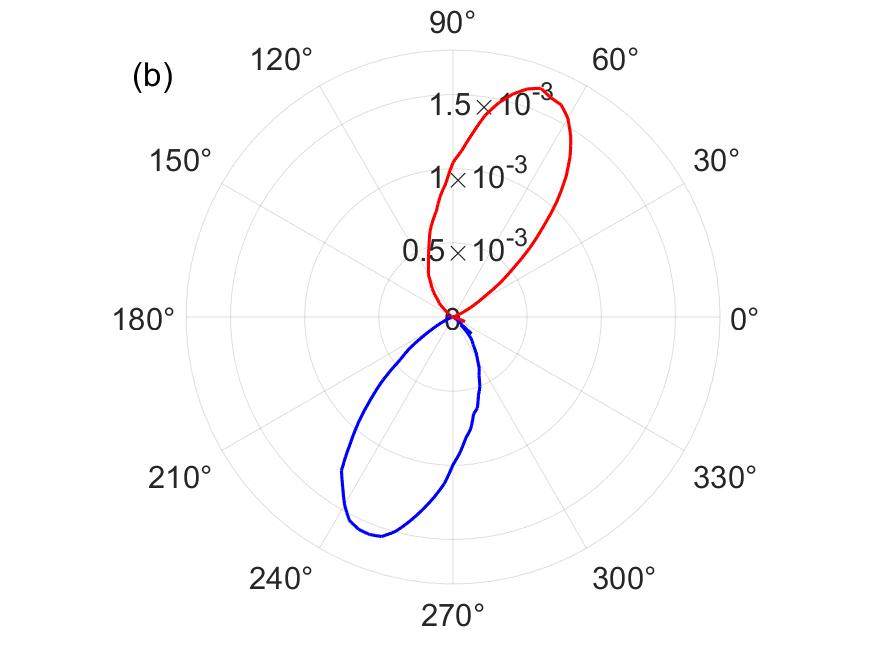}
    \end{subfigure}

    \caption{\textbf{Nonlinear Contribution of 1T$^\prime$ \WSe{2}} (a) THz amplitude as a function of azimuthal angle $\phi$ for the Bi$_2$Se$_3$/\WSe{2}(2ML)/Co heterostructure showing the acquired nonlinear behaviour (black), the TI contribution (grey dash) and the extracted contribution for \WSe{2} only (orange). (b) Polar plot of \WSe{2} emission  showing only a two lobe patten as expected for the 1T$^\prime$ polymorph - Table~\ref{tab_layer}.}
    
    \label{fig_THz_extract}
\end{figure}

The 2 ML nonlinear form can be further analysed to extract the exact contribution of the \WSe{2} layers by removing the contribution from the TI. Figure~\ref{fig_THz_extract}(a) shows the THz emission of the 2 ML sample on a cartesian plot. Here, removing from the raw data the nonlinear contribution from the \BISE~layers as $\cos(3\phi)$ (grey dashed curve), results in the nonlinear contribution of \WSe{2} (orange curve). The data is plotted in polar coordinates in Figure~\ref{fig_THz_extract}(b), that shows a clear two fold geometry as expected for a 1T$^\prime$ polymorph of \WSe{2} (see Table~\ref{tab_layer}). Note that although 1T$^\prime$ WSe$_2$ likely grows in three domains, oriented every 120$^\circ$ on Bi$_2$Se$_3$, the observed 2-fold symmetry demonstrates that one domain orientation dominates. This preferential domain orientation is due to the vicinal character of the sapphire surface that appears to favor one domain orientation~\cite{Qin24_3R_epitaxy}. Indeed the miscut axis has been found along $\phi\approx 150$° by x-ray diffraction which is orthogonal  to the anisotropy axis experimentally observed in the nonlinear THz component of Figure ~\ref{fig:nonmag}(c).

Regarding the magnetic contribution of the THz signal (Figure~\ref{fig:mag}), this can be correlated to the efficiency of the spin injection and spin-charge conversion processes. First, the decrease of the magnetic contribution with \WSe{2} thickness is a result of the formation of an energy barrier~\cite{cheng_far_2019}, which prevents an efficient spin-injection from Co to Bi$_2$Se$_3$ where interfacial conversion occur. For 4 MLs \WSe{2}, the signal from the magnetic component stops decreasing, suggesting that the emission comes from the self emission of the ferromagnetic layer ~\cite{zhang_ultrafast_2020} and that no spin current is injected through \WSe2 layers. (Although the magnetic contributions reduce as \WSe{2} acts as a barrier, this can be overcome through the growth of 2D ferromagnets on top of the TI, followed by the TMD growth such that both the TMD and TI emission can contribute additionally). Secondly, Figure~\ref{fig:mag} also shows that the magnetic contribution is not entirely isotropic for 1 ML and 2 ML of \WSe{2}, which is not the case of the sample without \WSe{2} layers or for the 3 ML and 4 ML \WSe{2} samples (see Supp. Info.), and suggests another magnetic phenomena in the measured THz field. This is highlighted by subtracting the mean value of the THz field from the magnetic contribution, as shown in Figure~\ref{fig:mag}, with the colours showing positive (red) and negative (blue) values. As can be observed in the 1 ML and 2 ML \WSe{2} structures, the symmetry appears to resemble the form observed for the 2 ML nonlinear contribution but with opposite phase and without a contribution of \BISE~(Figure~\ref{fig_THz_extract}). This suggests that the form is related to the 1T$^\prime$ phase of \WSe{2} and its high anisotropic nature but of magnetic origin. The treatment of the data through Equation~\ref{eq:components} of the magnetic component removes the intrinsic part of the nonlinearity. With the constant part removed, this leaves only the magnetic nonlinearity $\chi^{(2)}_{c}$. As the form is exactly the same of intrinsic nonlinearity $\chi^{(2)}_{i}$ and that \WSe{2} is not itself magnetic, this suggests that the magnetic properties are induced by the proximity of the TMD with the FM. Further, the opposite phase of the magnetic component suggests that a positive magnetic field gives a slightly smaller nonlinearity than negative fields i.e. $\chi^{(2)}_{c}$(+B) is negative and $\chi^{(2)}_{c}$(-B) is positive. The measured difference in magnetic and non magnetic nonlinearity is approximately an order of magnitude giving therefore a magnetic nonlinearity ten times smaller.  Interestingly the six fold symmetry from the 2H or 3R polymorph are not observed, highlighting that the nonlinearity of this polymorph is considerably smaller.

Note that the observed behavior could be considered as an extra spin-to-charge conversion process in the anisotropic 1T’ polymorph of \WSe{2}. Indeed, a considerable body of work has shown that with materials of reduced symmetry such as the 1T$^\prime$ investigated here demonstrate an anisotropic response in their spintronic properties (as well as optical, electrical, magnetic etc)~\cite{Tang21_1Tprime_review}. This has been concentrated on spin-to-charge conversion in the Te family (e.g WTe$_{2}$ and MoTe$_{2}$) that has a distorted crystal structure as in 1T$^\prime$ \WSe{2}. For example, junctions of WTe$_{2}$/FM have shown that charge conductivity, spin conductivity and spin-to-charge conversion is anisotropic and differs from the isotropic states of topological insulators~\cite{li_spin-momentum_2018,macneill_control_2017,Zhao20_CSC_WTe2,safeer_large_2019}. However, owing to the symmetry of the structure, a rotation of 180° of the sample would lead to exactly the same sign of the generated current. Therefore a $cos^2(\phi$) type dependence would be expected if there was an anisotropic spin-to-charge conversion contribution, rather than the $cos(\phi$) observed here. Although there could be other spin-to-charge phenomena, in this current work these appear quite weak, possibly owing to the the \WSe{2} acting as a spin barrier. Nonetheless the 1T$^\prime$ anisotropy clearly leads to magnetic dependent nonlinearity as a result of proximity effects with the FM.

\section*{Conclusions}

We have demonstrated the realisation of complex heterostructures of vdW materials and their polymorphs, showing high quality large area growth, where multiple physical phenomena can be combined over macroscopic surface areas. This is demonstrated in their giant nonlinearities, THz spintronics in topological insulators and magnetic proximity effects in reduced symmetry materials. The current work demonstrates the principle of this scaleable approach of combining different materials through growth techniques. A wide range of perspectives from this base of vdW heterotructure engineering will be possible with the first steps in more complex heterostructures with, for example, multiple periods to enhance the effects and combination with 2D ferromagnetics~\cite{chen_recentFM_2024, Guillet23_2DFM}. Longterm this work can be combined with other phenomena, such as the coupling with ferroelectricity, phononics and coherent current control to the effects already observed, which could be further manipulated either by a gate or through the excitation wavelength. This work opens up important prospects of new types of interfaces and devices on technological relevant levels, where the entire properties of the material can be artificially designed and controlled.

\begin{acknowledgments}
M. M. and A. W. contributed equally to this work. The authors acknowledge funding from European Union’s Horizon 2020 research and innovation program under grant agreement No 964735 (FET-OPEN EXTREME-IR) and the French National Research Agency (ANR) under ANR-22-CE30-0026 DYNTOP. This work was supported by a government grant managed by the ANR as part of the France 2030 investment plan from PEPR SPIN ANR-22-EXSP-0002. \\
This work was granted access to the HPC resources of MesoPSL financed by the Region Ile de France and the project EquipMeso (reference ANR-10-EQPX-29-01) of the programme Investissements d’Avenir supervised by the Agence Nationale pour la Recherche.
The authors acknowledge funding from the European Union's Horizon 2020 research and innovation program under grant agreement No 881603 (Graphene Flagship) and No 101099552 (HORIZON-EIC-2022-PATHFINDEROPEN-01-01 PLASNANO). The French National Research Agency (ANR) is acknowledged
for its support through the ESR/EQUIPEX+ ANR-21-ESRE-0025 2D-MAG project. The LANEF framework (No. ANR-10-LABX-0051) is acknowledged for its support with mutualized infrastructure.\\
\end{acknowledgments}


\section*{Competing interests}

The authors declare no conflict of interest.


\bibliography{TI-TMD-FM}


\end{document}